\documentclass[prd,aps,twocolumn,nofootinbib,superscriptaddress]{revtex4-1}

\usepackage{amsmath}
\usepackage{epsfig}
\usepackage{appendix}
\usepackage{subfigure}

\begin{document}

%\hfill{Preprint Number here}

\title{Multiple Cosmic Collisions and the Microwave Background Power Spectrum}

\author{Anthony Aguirre}
\email{aguirre@scipp.ucsc.edu}\affiliation{Department of Physics, University of California, 1156 High St., Santa Cruz, CA 95064, USA}\affiliation{Santa Cruz Institute for Particle Physics, Santa Cruz, CA 95064, USA} 

\author{Jonathan Kozaczuk}
\email{jkozaczu@ucsc.edu}\affiliation{Department of Physics, University of California, 1156 High St., Santa Cruz, CA 95064, USA}

\date{\today}

\begin{abstract}

Collisions between cosmic bubbles of different vacua are a generic feature of false vacuum eternal inflation scenarios.  While previous studies have focused on the consequences of a single collision event in an observer's past, we begin here an investigation of the more general scenario allowing for many ``mild" collisions intersecting our past light cone (and one another).  We discuss the general features of multiple collision scenarios and consider their impact on the cosmic microwave background (CMB) temperature power spectrum, treating the collisions perturbatively.  In a large class of models, one can approximate a multiple collision scenario as a superposition of individual collision events governed by nearly isotropic and scale-invariant distributions,  most appearing to take up less than half of the sky.  In this case, the shape of the expected CMB temperature spectrum maintains statistical isotropy and typically features a dramatic increase in power in the low multipoles relative to that of the best-fit $\Lambda$CDM model, which is in tension particularly with the observed quadrupole.  We argue that this predicted spectrum is largely model-independent and can be used to outline features of the underlying statistical distributions of colliding bubbles consistent with CMB temperature measurements.  %Upon first look, scenarios with many mild collisions may be compatible with observations of the CMB, although they tend to amplify the observed quadrupolar power deficit.

\end{abstract}

\maketitle

\section{Introduction}

The theory of cosmological inflation has been very successful in proving potential answers to several deep questions in early universe cosmology, as well as testable (and tested) cosmological predictions.  However, a crucial ``side effect" of many versions of inflation is that the exponential expansion tends to continue on eternally into the future, with only pockets of the universe ceasing to inflate locally.  This scenario is known as eternal inflation (see e.g. Refs~\cite{Aguirre:2007gy, Guth:2007ng, Linde:2007fr} for reviews).

Eternal inflation can be driven by several mechanisms.  In the present study, we concern ourselves with the ``false vacuum" variety, in which inflation occurs as a result of an inflaton (usually taken to be an effective scalar field, $\varphi$) being trapped in a metastable false vacuum.  Inflation ends locally when the field transitions to a nearby ``truer" vacuum on the potential.  If the probability per unit four-volume of such a transition, $\lambda$, is small compared to the Hubble scale of the inflating false vacuum $H_F^{4}$ (as typically assumed), then false vacuum inflation is eternal. For many potentials, the dominant transition mechanism is bubble nucleation \cite{CDL}, which can quite naturally produce open FRW universes much like our own.  In this study, we will work under the assumption that we live in one such universe, dubbed ``the observation bubble".

A fascinating consequence of false vacuum eternal inflation is that other bubbles inevitably form in the four-volume to the past of the observation bubble wall.  These bubbles collide with the observation bubble wall, potentially leading to detectable signatures of ``cosmic bubbles" outside of our own, provided that an observer can exist to the future of the collision.   Much recent work has been dedicated to the study of such relics~\cite{Garriga:2006hw,  Aguirre:2007an, Aguirre:2007wm, Aguirre:2008wy, Chang:2007eq, Chang:2008gj, Freivogel:2009it, Larjo:2009mt, Czech:2010rg, Easther:2009ft, Giblin:2010bd, Johnson:2010bn, Kleban:2011yc, Gobbetti:2012yq, Feeney:2010dd, McEwen:2012uk, Johnson:2011wt, Hwang:2012pj} (for a general review, see e.g. Ref.~ \cite{Aguirre:2009ug}).  In particular, it has been shown that if a collision is ``mild" enough, it can potentially be compatible with the formation and survival of observers in the post-collision region \cite{Aguirre:2007wm,Aguirre:2008wy} while producing testable predictions for observations of the Cosmic Microwave Background (CMB)\cite{Chang:2008gj,Kleban:2011yc,Feeney:2010dd}, bulk galactic flow \cite{Larjo:2009mt}, and possibly other cosmological observables (see e.g. Ref.~\cite{Kleban:2011pg} for a discussion).

While the observation bubble wall undergoes a divergent number of collisions, whether or not these collisions lie within the observer's past lightcone at decoupling depends on several factors, such as the position of the observer and the cosmology inside the observation bubble,  which we review in more detail in Sec.~\ref{sec:distributions}.  Previous studies have focused on the outcome of a single observable collision, or a few collisions whose affected post-collision regions do not intersect.  In these cases, the azimuthal symmetry of the collisions' perturbations to the observation bubble spacetime is maintained and one expects to see an azimuthally symmetric disk on the CMB sky corresponding to each collision \cite{Chang:2008gj,Kleban:2011yc}.  Recent studies \cite{Feeney:2010dd,McEwen:2012uk} have searched for these disk-like patterns in the WMAP7 CMB temperature data and have found a few candidates for collision events.

As we will see in Sec.~\ref{sec:distributions}, however, for arbitrary choices of the various parameters, the number of observable bubbles $N$ can potentially range from zero to very large numbers, and there is no particular reason to expect $\sim$ one observable collision rather than none or many.  When $N\gg1$, the picture differs from that of previous studies in several respects: \begin{enumerate} \item The future light cones of the different collisions will tend to intersect one another.  Thus, the effects arising from various collisions can potentially interact.  \item One must address the possibility that some bubbles may in fact collide with each other before impacting the observation bubble wall, leading to a qualitatively different collision scenario than before. \item Since any given part of an observer's LSS will have been affected by several different collisions, the azimuthal symmetry of an isolated collision is no longer evident. \item With a large number of bubble collisions, one may be able to make meaningful statistical predictions about what a given observer can expect to see. \end{enumerate}  The purpose of this study is to explore these issues and take the first step towards assessing how many-bubble scenarios can be constrained by current and future observations.  As a first look, we will focus primarily on the effect of the collisions on the CMB temperature, deferring an analysis of the other relevant observables to future study. 

Clearly if we are to be living in such a large-$N$ scenario, the effect of each collision must in some sense be ``small" so as not to disrupt the subsequent cosmological evolution in the observation bubble.  As we will discuss in Sec.~\ref{sec:mult_prof}, the strength of a collision is determined by the position of the nucleation center of the colliding bubble in the exterior de Sitter (dS) space, as well as the details of the underlying potential, which are unknown.  Thus, in analyzing scenarios with $N>1$, we will simply start with the assumption that an observer exists with a large number of collisions intersecting the visible portion of the last scattering surface and study what might be observed.  This is in keeping with previous work on the CMB temperature profile for single collisions.  We do not consider scenarios in which some collisions might prevent inflation to the future of the collision, such as those in small-field type potentials (see the discussion in Ref.~\cite{Aguirre:2009ug}).  When discussing an observer's expectations, we therefore mean the expected value considering only mild collision scenarios. 

%The point is, when we talk about "expected profile", we are not factoring in how likely it is that an observer exists or not; say "given that an observer has formed in a multiple collision scenario, what would he expect to observe?"

The current study is organized as follows: in Secs.~\ref{sec:distributions}-\ref{sec:mult_prof} we set up the various parts of the problem and argue that many scenarios with multiple bubble collisions can be considered as a superposition of individual impacts, each appearing to take up less than half of the observer's sky.  In Sec.~\ref{sec:power_spect} we compute the expected signal in the CMB temperature power spectrum.  Sec.~\ref{sec:conclusion} contains our conclusions.

\section{Bubbles, Collisions, and Observers} \label{sec:distributions}

\subsection{The Setup}
%%%%%%%%%%%%%%%%%%%%%%%%%%%%%%%%%%%%%%%%%%%%%%%%%%%
\begin{figure}[!t]
\mbox{\includegraphics[width=0.3\textwidth,clip]{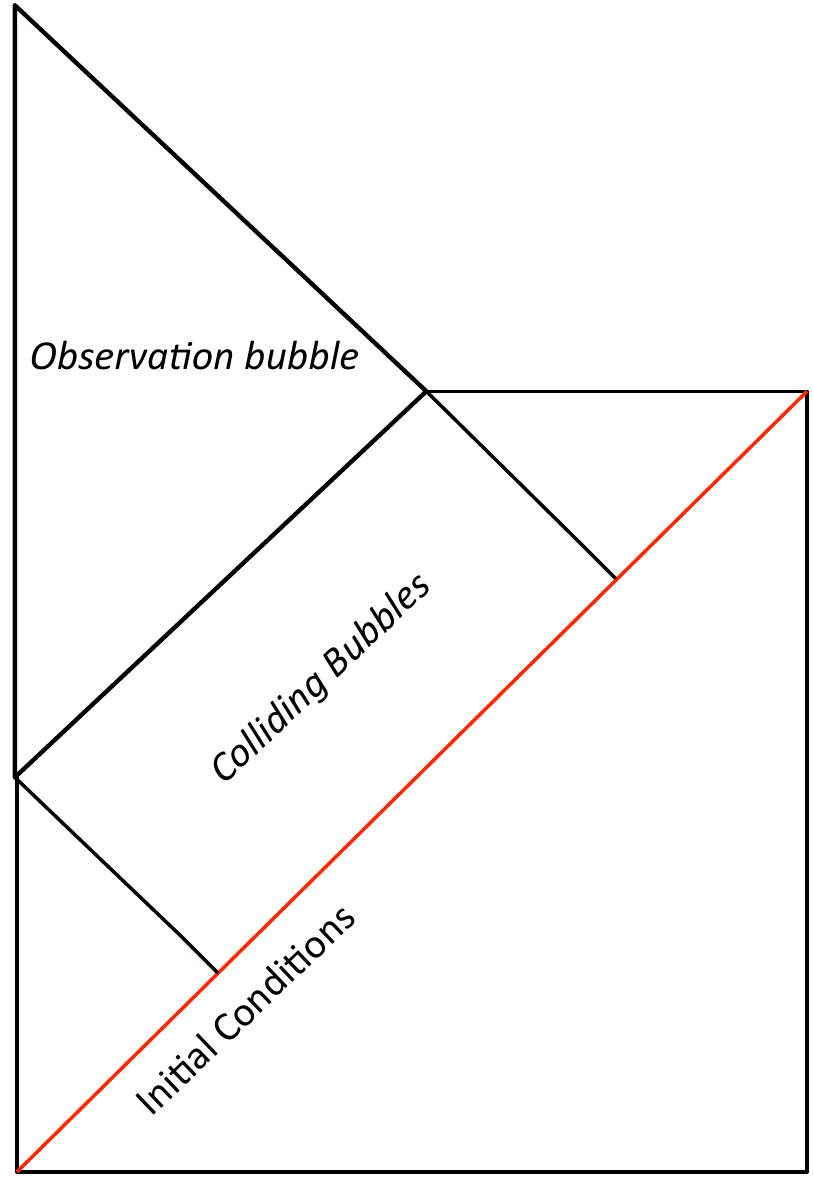}}
\caption{\label{fig:conformal_1}\small Conformal slice of the basic bubble collision setup in the so-called ``steady state frame".  The observation bubble shown is pure dS with $H_F/H_I\gg1$.  The region out of which colliding bubbles is labelled as such, and the red line demarcates the initial value surface.}
\end{figure}
%%%%%%%%%%%%%%%%%%%%%%%%%%%%%%%%%%%%%%%%%%%%%%%%%%%

We begin with a false vacuum, which we take to be dS space with associated Hubble constant $H_F$.  We can describe the false vacuum dS as a hyperboloid $\eta_{\mu\nu}X^{\mu}X^{\nu}=H_F^{-2}$ embedded in a 5-dimensional Minkowski space with coordinates $X^{\mu}$ and metric $\eta_{\mu \nu}$, where $\mu,\nu=0,\ldots,4$.    The most useful coordinatization of dS for our purposes will be the so-called ``global slicing", with coordinates $(T,\eta,\theta,\phi)$ defined in the embedding space by \begin{equation} \begin{aligned} \label{eq:embedding} X_0&=H_F^{-1}\tan T \\ X_i&=H_F^{-1}\frac{\sin\eta}{\cos T} \omega_i \\ X_4&=H_F^{-1}\frac{\cos \eta}{\cos T}, \end{aligned} \end{equation} where $-\pi/2\leq T\leq \pi/2$, $0\leq \eta \leq \pi$, and $\omega_i=(\cos\theta, \sin \theta \cos\phi, \sin \theta \sin \phi)$.  This induces the metric 
\begin{equation} 
ds^2 = \frac{1}{H_F^2 \cos^2 T}\left(-dT^2 + d\eta^2+\sin^2 \eta\hspace{1mm}d\Omega_2^2\right),
\end{equation}  
which covers the entire dS manifold.  The 4-volume element in these coordinates is 
\begin{equation}
\label{eq:4vol} dV_4= H_F^{-4} \frac{\sin\eta}{\cos^4 T}dT d\eta \hspace{1mm} d\Omega_2 .
\end{equation}

The basic setup we consider is the same as that of Refs.~\cite{Aguirre:2007an, Aguirre:2007wm, Aguirre:2008wy, Aguirre:2009ug}, to which we refer the reader for further details.  A small, thin-walled observation bubble is nucleated out of the false vacuum via the Coleman-deLuccia (CDL) process~\cite{CDL} and subsequently expands with a trajectory which we approximate as null.\footnote{We note that, while this small-bubble, thin-wall limit~\cite{Aguirre:2007an} is convenient and used here for both collision and observation bubbes, it is not necessarily natural and future work is needed to go beyond this approximation.} The observation bubble starts off curvature dominated and undergoes a period of subsequent slow-roll inflation, which we model as dS space with Hubble constant $H_I$.  Throughout this study, we assume $H_F/H_I\gg1$ except where noted; this seems natural, as $H_I<H_F$ for tunneling to occur, and there is no obvious reason why the scale of inflation in the observation bubble should be similar to the scale associated with the false vacuum.  After inflation, we assume a standard $\Lambda$CDM cosmology, although we typically ignore matter domination and the possibility of late-time accelerated expansion in our calculations, which should induce small corrections.  This scenario is sketched in Fig.~\ref{fig:conformal_1} for a conformal slice of $\theta, \phi=$const. (Note that as drawn, the observation bubble is pure dS with $H_F/H_I\gg1$).  In this diagram the bottom (top) boundary  of the false vacuum corresponds to $T=-\pi/2$ $(\pi/2)$, while $\eta$ ranges from $0$ to $\pi$ from left to right.  

The observation bubble interior spacetime is given by the analytic continuation of the CDL instanton \cite{CDL}, which, in our approximation, yields an open FRW cosmology inside the null cone emanating from the nucleation center, with metric \begin{equation} ds^2=-d\tau^2+a^2(\tau)\left(d\xi^2+\sinh^2\xi d\Omega_2^2\right) .\end{equation}  Spacelike surfaces of constant $\tau$ are 3-hyperboloids, $H_3$, reflecting the $SO(3,1)$-invariance of the CDL instanton.  The observation bubble and false vacuum dS spacetimes can be joined together by gluing across the bubble wall, as described in Ref.~\cite{Aguirre:2007an}.
%The observation bubble interior spacetime is given by the analytic continuation of the CDL instanton \cite{CDL}, which, in our approximation, yields an open FRW cosmology inside the null cone emanating from the nucleation center.  The embedding \begin{equation} \begin{aligned} X_0&=a(\tau)\cosh \xi \\ X_i&=a(\tau)\sinh \xi \omega_i \\ X_4&=f(\tau) \end{aligned}\end{equation} with $0\leq \xi<\infty$, $0<\tau<\infty$ and where $f(\tau)$ solves $f'^2 (\tau)=a'^2(\tau)-1$, induces the metric \begin{equation} ds^2=-d\tau^2+a^2(\tau)\left(d\xi^2+\sinh^2\xi d\Omega_2^2\right) \end{equation} in the observation bubble.  Setting $a(\tau)=H_I^{-1}\sinh(H_I \tau)$ yields the open slicing of dS.  Spacelike surfaces of constant $\tau$ are 3-hyperboloids, $H_3$, reflecting the $SO(3,1)$-invariance of the CDL instanton.  The observation bubble and false vacuum dS spacetimes can be joined together by gluing across the bubble wall, as described in Ref.~\cite{Aguirre:2007an}.

Along with the observation bubble, other bubbles can also nucleate out of the false vacuum with associated probability per unit 4-volume $\lambda_i$, where $i$ represents the bubble's corresponding vacuum.  The region of dS that can nucleate colliding bubbles is labelled in Fig.~\ref{fig:conformal_1} in the ``steady state" frame in which the distribution of bubbles is statistically independent of position and time, and the conformal diagram is independent of angle.  Bubbles nucleated above the top boundary of this region never intersect the observation bubble wall, while the lower boundary ensures that the observation bubble was indeed nucleated out of the false vacuum and not some other bubble.\footnote{Throughout this study we work in the approximation in which no bubbles can nucleate within another bubble.  This approximation is discussed in more detail in Sec.~\ref{sec:mult_prof}.}  Bubbles formed to the right of the red boundary will have future light cones that encompass future infinity of the false vacuum dS space, and so as in previous studies we postulate a ``no bubble" initial condition surface as shown.    

While the diagram in Fig.~\ref{fig:conformal_1} depicts the situation in the steady-state frame, there are other reference frames that will be useful to consider.  One can define a frame in which the observer is at the center of the observation bubble $\xi=0$ (the ``observation frame"), as well different ``collision frames" in which the observation bubble and a given colliding bubble nucleate at the same global slicing time $T=0$.  To boost between these different frames in both the observation bubble and exterior dS, one can apply the appropriate Lorentz transformation in the embedding space \footnote{One should note that in general once there are collisions it will take more dimensions to embed the spacetime and boost.}. As discussed in detail in Ref.~\cite{Aguirre:2007an}, in the observation frame the distance from the observation bubble wall to the initial value surface is angle-dependent.  This fact plays an important role when computing the distribution of bubbles an observer expects to see, as we will discuss in the following subsection.  

%We can define the initial value surface as $X_0+X_4=0$ (where $X_4=const$ describes the bubble wall), in which case the initial value surface is given by \begin{equation} T=\eta-\pi/2 \end{equation} in the steady state frame as shown in Fig.~\ref{fig:conformal_1}.  Boosting to the observation frame via $\gamma=1/\sqrt{1-v^2}$, the initial value surface transforms as \begin{equation} \label{eq:IVS} \sin T'=-\left(\frac{\cos\eta'}{\gamma}+v \sin \eta' \cos\theta ' \right). \end{equation} This fact plays an important role when computing the distribution of bubbles an observer expects to see, as we will discuss in the following subsection.  
Since a collision preserves an $SO(2,1)$ symmetry, it can be fully described in 1+1 dimensions, with each point representing a 2-hyperboloid with line element $dH_2$.  The dS false vacuum can be conveniently foliated according to this symmetry, yielding the metric
\begin{equation} ds^2= -(1+H^2 z^2)^{-1}dz^2+(1+H^2 z^2)dx^2+z^2dH_2^2. 
\end{equation} 
We can then place the observation and colliding bubble nucleation centers along the $x$-axis, and describe the kinematics of the collision by $(x_c, z_c)$, the nucleation center of the colliding bubble (the observation bubble is at the origin).  In the collision frame, this simplifies further: since both bubbles, by construction, nucleate at $z=0$, all of the kinematics are described by $z_c$, which corresponds uniquely to the de Sitter-invariant distance ($\eta_C$ in the collision frame) separating the bubbles.  This quantity is responsible for determining the strength of a given collision, as we will discuss in Sec.~\ref{sec:mult_prof}.      
%It will also be useful to define the quantity $z_c$ which, in a collision frame, corresponds uniquely to the de Sitter-invariant separation of the observation and colliding bubbles, $\eta_C$ (see Ref~\cite{Aguirre:2007wm} for a more detailed discussion).  For observers with $\xi_0=0$, $z_c$ is defined for a colliding bubble nucleated at $(\eta_n,T_n)$ as \begin{equation} \label{eq:z_c} z_c = H_F^{-1}\frac{\cos T_n - \cos\eta_n}{\sqrt{\sin^2\eta_n - \sin^2 T_n}}. \end{equation}  This quantity (or equivalently the dS-invariant distance) is responsible for determining the strength of a given collision, as we will discuss in Sec.~\ref{sec:mult_prof}.

\subsection{Probabilities And Observer Expectations}
\label{subsec:prob}

Ascertaining the general features of a multiple collision scenario requires information about the various parameters that enter into the theory -- in particular, the expected number of observable bubbles intersecting the observer's last scattering surface, as well as the distributions of the angular sizes and strengths of the collisions that would enter into e.g. the CMB temperature profile.  This section reviews the extensive recent progress made in understanding these distributions, focusing on what is relevant to the task at hand and offering some novel insights.  We refer the reader to Ref.~\cite{Aguirre:2009ug} for a good summary and for details of many of the calculations discussed here.

We begin in the observation frame, in which $\xi_0=0$.  Under our assumptions,  the number of bubbles intersecting part or all of the observer's last scattering surface is
 \begin{equation} 
 \label{eq:N} 
 N=\lambda V^{obs}_4, 
 \end{equation} 
 where $V^{obs}_4$ is the 4-volume in the false vacuum dS spacetime out of which such a bubble can nucleate, computed by integrating the 4-volume element Eq.~\ref{eq:4vol} over the relevant region.  There are two regions in the exterior dS in which $dV_4$ can potentially greatly exceed its natural scale of $H_F^{-4}$, and hence contribute the most to $V_4^{\rm obs}$ and the total count of bubbles.  

First, for observers with $\xi_0\rightarrow \infty$, there is a large contribution to $N$ from portions of $V_4^{\rm obs}$ near past infinity of the false vacuum.  This results in an anisotropic peak in the distribution of bubbles in the direction towards the initial value surface (defined as $\theta=0$), which can be thought of as the time dilation of the observer's time relative to the ``steady state" cosmological time.  This anisotropy was first recognized in Ref.~\cite{Garriga:2006hw} and dubbed the ``persistence of memory".  Bubbles nucleated in the region near past infinity will cover essentially the full sky and enter the observer's past light cone at small $\tau$ and were thus dubbed ``early time" collisions in Ref.~\cite{Aguirre:2007an}.  As discussed in previous work these collisions will be hard or impossible to observe, and we will exclude them from our analysis, while noting that they still comprise something of an open problem as to precisely how this infinity should be regulated.

A second sizable contribution to $V_4^{\rm obs}$ results for large $H_F/H_I$ from bubbles appearing on the conformal diagram near future infinity of the false vacuum.  This enters the observer's past lightcone when the ``hat" in the conformal diagram (Fig.~\ref{fig:conformal_1}) extends high enough, which in turn happens for $H_F \gg H_I.$
This effect is largely independent of $\xi_0$ and corresponds to ``late time" collisions of all angular scales, and nearly uniformly distributed in $\cos \theta$.  

%%%%%%%%%%%%%%%%%%%%%%%%%%%%%%%%%%%%%%%%%%%%%%%%%%%
\begin{figure*}[!t]
\mbox{\hspace*{-1.2cm}\includegraphics[width=.9\textwidth,clip]{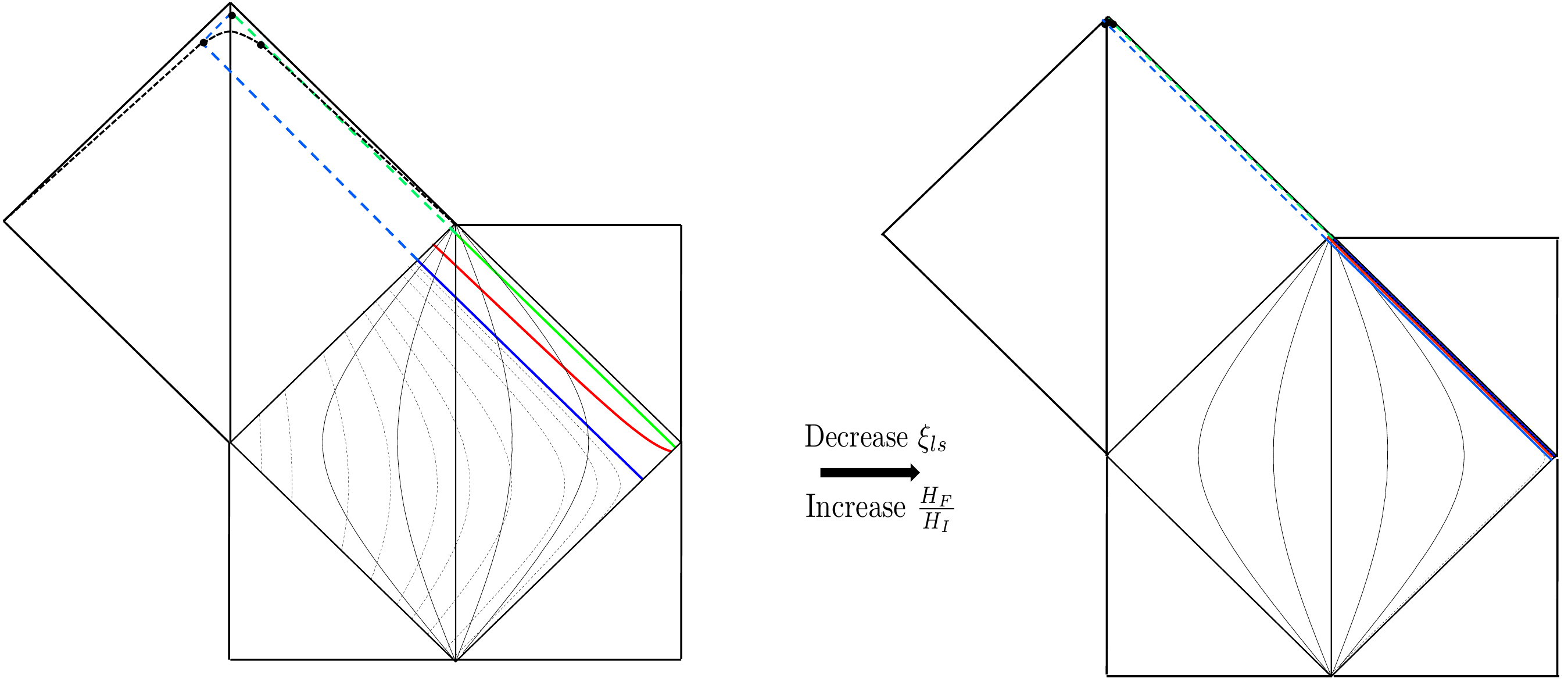}}
\caption{\label{fig:conformal_2}\small Foliation of a $\theta=0$ conformal slice of the external de Sitter spacetime for an observer at $\xi_0=0$ with lines of constant $\rho$, $z_c$ for the (unrealistic) values $\xi_{ls}=1$ and $H_F/H_I=10$ on the left, and for $\xi_{ls}=.05$ and $H_F/H_I=100$ on the right.  The observer's LSS is sketched by the dashed black line in the observation bubble (this curve will be compressed up towards the tip of the observation bubble for the case shown on the right).  Solid black curves correspond to different values of $z_c$, with the vertical line in the center corresponding to $z_c=H_F^{-1}$ in both cases.  The dotted curves represent different values of $\rho$, with $\rho=-1$, $0$, $1$ denoted by the solid green, red, and blue curves, respectively.  The effect of decreasing $\xi_{ls}$ and increasing $H_F/H_I$ is shown on the right.  The same range of $\rho$ values are plotted, only now all constant-$\rho$ curves depicted on the left are compressed into the thin colored region.}
\end{figure*}
%%%%%%%%%%%%%%%%%%%%%%%%%%%%%%%%%%%%%%%%%%%%%%%%%%%

Note that many late-time collisions will result in effects on scales greater than the observer's full sky.  To see why this is the case, we can coordinatize the exterior dS in terms of angular scale and $z_c$.  This was done in Ref.~\cite{Aguirre:2009ug} which showed that in the observation frame, each bubble intersecting part of the observer's LSS (but not encompassing the entire surface) with solid angular scale $\psi$ and dS-invariant distance corresponding to $z_c$ can be mapped uniquely to a nucleation point $(\eta_n, T_n)$ in the exterior dS.  A straightforward generalization of Appendix D of Ref.~\cite{Aguirre:2009ug} allows us to exchange the angular coordinate $\psi$ for one describing the size of greater than full-sky bubbles as well; this quantity, which we call $\rho$, is defined via \begin{equation} \label{eq:l} \begin{aligned} \rho\equiv&\frac{-1}{\sinh H_I \tau_{ls}\sinh\xi_{ls}}\left[\frac{\left(1+H_I^2z_c^2\right)}{2H_Iz_c\gamma}\right. \\ & \left. -\frac{\left(1-H_I^2z_c^2\right)\cosh H_I\tau_{ls}}{2H_Iz_c\gamma}  + v\sinh H_I\tau_{ls}\cosh\xi_ls \right]  \end{aligned}  \end{equation} where $\xi_{ls}=\int_{\tau_{ls}}^{\tau_0}d\tau/a(\tau)$ is the comoving radius out to which the observer can see on the LSS \footnote{Measurements of the curvature density \cite{Aguirre:2009ug,WMAP2} dictate that $\xi_{ls}\sim 2 \sqrt{\Omega_c}\lesssim 0.18$.  In what follows, we thus assume $\xi_{ls}$ is small, typically taking $\xi_{ls}\approx 0.05$ as an illustrative value.}.  For $-1\leq \rho\leq 1$, $\rho$ corresponds exactly to $-\cos\psi/2$.  The coordinate $\rho$ effectively measures the distance out on the observer's last scattering surface to its intersection with the future light cone of a given collision in the observation frame.  Values $\rho<-1$ correspond to late-time collisions which do not intersect any of the observer's LSS, while $\rho>1$ corresponds to a greater-than-full-sky collision.  

The foliation\footnote{For $\rho>1$ there can be two points in the false vacuum corresponding to the same value of $(z_c, \rho)$ as seen on the LHS of Fig.~\ref{fig:conformal_2}.  However, we are most often interested in $\left|\rho\right|\leq 1$, for which the foliation is one-to-one.} of the exterior dS in terms of $(\rho, z_c)$ is shown in Fig.~\ref{fig:conformal_2} for $H_F/H_I=10$, $\xi_{\rm ls}=1$ on the left, and for $H_F/H_I=100$, $\xi_{ls}=0.05$ on the right.  The colored solid lines correspond to $\rho$ values such that $\left| \rho\right| \leq1$, i.e. nucleation centers of bubbles with angular scales $0\leq \psi\leq 2 \pi$.  The ($\theta$-dependent) initial value surface is not shown.  Notice that for small values of $\xi_{ls}$ (as required by observation) and $H_F/H_I\gg1$, bubbles of virtually all angular scales (including $\rho>1$) receive a large 4-volume contribution from points near future infinity of the exterior dS space.  This is true regardless of the position of the initial value surface and hence of the angles $\theta$, $\phi$.  Thus, neglecting early-time bubbles and for values of $H_F/H_I$, $\xi_{ls}$ we are interested in, an observer should expect to see a nearly scale-invariant, isotropic distribution of  collision bubbles, regardless of his or her position in the observation bubble.  (This typically includes many greater-than-full-sky late-time bubbles, as is evident from Fig.~\ref{fig:conformal_2}.)  As the relevant constant-$\rho$ curves are compressed into a very thin region in this case, from Fig.~\ref{fig:conformal_2} one would expect the scale-invariance to hold true independent of $z_c$, although the distribution of bubbles itself will have $z_c$-dependence.  

Explicitly calculating the distribution of bubbles over angular size by integrating the 4-volume element over the constant-$\rho$ contours in the relevant region yields \begin{equation} \frac{dN}{d\rho d\Omega_2}\simeq \lambda H_F^{-4} \left(\frac{H_F}{H_I}\right)^2\xi_{ls} \end{equation} for $\xi_{ls}\rightarrow 0$, which is indeed flat in $\rho$.  This distribution was first obtained by Ref.~\cite{Freivogel:2009it} and dubbed the ``disintegration of the persistence of memory" due to its isotropy. 

% [[Is another way to say this that as we follow a line of constnat $z_c$ through the (compressed) lines of constant $l$, the integration volume factor from eq. 3 stays constant? ]] As the relevant constant-$\rho$ curves are compressed into a very thin region in this case, from Fig.~\ref{fig:conformal_2} one would expect the scale-invariance to hold true independent of $z_c$, although the distribution of bubbles itself will have $z_c$-dependence.  Explicitly calculating the distribution of bubbles over angular size by integrating the 4-volume element over the constant-$\rho$ contours in the relevant region yields \begin{equation} \frac{dN}{d\rho d\Omega_2}\simeq \lambda H_F^{-4} \left(\frac{H_F}{H_I}\right)^2\xi_{ls} \end{equation} for $\xi_{ls}\rightarrow 0$ which is indeed flat in $\rho$.  

Integrating the 4-volume element over all values of $\rho\geq-1$ and $z_c$ in the region depicted in Fig.~\ref{fig:conformal_1} yields the expected total number of bubbles intersecting the observer's LSS, given by \begin{equation} \label{eq:N_tot} N \simeq \frac{4\pi \lambda}{3 H_F^4}\left(\frac{H_F^2}{H_I^2}\right) \xi_{0} \end{equation} for observers at  $\xi_0\rightarrow \infty$ \footnote{Virtually all observers, except a set of measure zero, are expected to fall into this category.}.  Limiting the integration to $-1\leq \rho\leq1$ removes greater-than-full-sky bubbles from the count and yields \begin{equation} \label{eq:Nobs} N_{\rm obs} \simeq \frac{16 \pi\lambda }{3 H_F^4}\left(\frac{H_F^2}{H_I^2}\right)\sqrt{\Omega_c}.\end{equation} 

Comparing $N$ to $N_{\rm obs}$ we see explicitly that for each late time sub-horizon bubble, there will be $\sim \xi_0 \Omega_c^{-1/2} \gg 1$ full-sky bubbles to account for when computing the effects of bubble collisions.  As we will argue in the following sections, however, collisions that encompass the entire visible portion of the LSS will result in adiabatic superhorizon perturbations to the gravitational potential at the LSS and hence not contribute significantly to the CMB temperature spectrum.  

From Eq.~\ref{eq:Nobs}, there are many possible potentials and nucleation rates giving rise to any particular value of $N_{\rm obs}$.  Our approach will be to treat $N_{\rm obs}$ as a free parameter when computing the expected CMB temperature spectrum and we will not attempt to address how feasible or likely scenarios with a given $N_{\rm obs}$ might be, aside from the following general comments.   First, from Eq.~\ref{eq:Nobs}, even for $\lambda H_F^{-4}\ll1$, one can still expect $N_{\rm obs}\gg1$ as long as $H_F/H_I$ is large enough to compensate for the small nucleation rate.  Second, it is clear from Eq.~\ref{eq:Nobs} that $N_{\rm obs}=1$ is likely a very special case, requiring the precise balance between features of the potential through $H_F$, $H_I$, $\Omega_c$, and $\lambda$ as mentioned previously.

Finally, one can also compute the distribution of colliding bubbles with respect to $z_c$.  This calculation was carried out by the authors of Refs.~\cite{Aguirre:2007wm, Aguirre:2009ug} who found that the distribution $dN/dz_c d\Omega_2$ is peaked sharply around $z_c\sim H_F^{-1}$ and falls off as $\sim z_c^{-3}$ for large $z_c$.

\section{The Profile: Single Bubble Case}\label{sec:profile}

Calculating the CMB temperature spectrum resulting from a multiple collision scenario requires an understanding of the temperature profile associated with a single bubble collision, which we discuss here, as well as how to combine many such collisions, which is discussed in the next section.  For more detailed derivations of the single-collision temperature profile we refer the reader to Refs.~\cite{Chang:2008gj, Kleban:2011pg, Kleban:2011yc}.  

\subsection{The Single-Bubble Profile As A Small Potential Perturbation} \label{subsec:single}

%\subsubsection{Symmetry Consideration}

As noted above and in previous studies, the symmetry of a bubble collision leads to an azimuthally symmetric perturbation on an observer's LSS, filling a disk that constitutes the triple-intersection of the last-scattering surface, the observer's past lightcone, and the future light cone of the bubble collision.  Within this disk, given the necessary slow-roll inflation in the observation bubble and the lack of large anomalies on the actual CMB, the effect of the collision is necessarily small, and can be described as a perturbation to the gravitational potential $\Phi(\vec{x})$ at the reheating time $\tau_{rh}$.  In conformal Newtonian gauge and retaining only the term linear in slow roll (see Ref.~\cite{Kleban:2011yc}), the gravitational potential resulting from the inflaton perturbation is given by \begin{equation}\label{eq:pert1} \Phi(\tau_{rh},\vec{x})=\bar{\alpha} \xi_{rh}(\cos\theta-\cos\frac{\psi}{2})\Theta(\cos\theta-\cos\frac{\psi}{2})\end{equation}  where coordinates are chosen so that the collision is centered around $\theta,\phi=0$ with $\theta=\psi/2$ the causal boundary of the collision, $\Theta(x)$ is the Heaviside step function, and $\xi_{rh}$ is the comoving distance at which the observer's past light cone intersects the reheating surface.  The parameter $\bar{\alpha}$ contains all the information about the inherent strength of the perturbation, which depends on the kinematics and the shape of the inflationary potential.

Equivalently, the gravitational potential Eq.~\ref{eq:pert1} can be exchanged for the gauge-invariant curvature perturbation, which at late times during inflation is simply $\zeta \propto \Phi$; the proportionality constant here depends on the inflationary potential and $\bar{\alpha}$.  Once $\zeta$ is known, one can use the Sachs-Wolfe approximation \cite{Sachs:1967er} $\Delta T /T \propto \zeta(x,\tau_{ls})$ to determine the CMB temperature spectrum resulting from the collision.  Here the factors relating $\zeta$ to $\Delta T/T$ depend only on the background cosmology.  Absorbing $\bar{\alpha}$ and the other relevant parameters fixed by the potential and intra-bubble cosmology into the parameter $\alpha$, and neglecting (temporarily) the inflationary temperature fluctuations, as well as the evolution of the perturbation between reheating and decoupling,\footnote{Technically one should take into account the evolution of $\zeta$ between reheating and decoupling by using the appropriate transfer functions.  This was carried out in Ref.~\cite{Kleban:2011yc} which found that the results are in good agreement with Eq.~\ref{eq:prof}. (This is expected, since for $H_F/H_I\gg1$ the conformal time elapsed between reheating and decoupling is negligible.)  For the remainder of this study we thus approximate the reheating time $\tau_{rh}\approx\tau_{ls}$.} we arrive at the approximate CMB temperature profile for a single collision \begin{equation}\label{eq:prof}  T(\theta,\phi)=T_0\left[1+\alpha \xi_{ls}(\mu-\mu_0)\Theta(\mu-\mu_0)\right] \end{equation} where $\mu=\cos\theta$, $\mu_0=\cos \psi/2$ for a collision centered at the north pole and $T_0$ is the average temperature of the unaffected portion of the LSS.

In sum, the form of the perturbation in the current approximation depends only on the residual $SO(2,1)$ symmetry of the post-collision spacetime, and all the information about the microphysics of the collision, including the details of the underlying potential and background cosmology, is encoded into a single parameter $\alpha$ describing the inherent ``brightness" of the collision on the CMB.  This fact will simplify the calculation of the expected CMB temperature spectrum for a multiple collision scenario in Sec.~\ref{sec:power_spect}. 

%It is important to note that the profile Eq.~\ref{eq:prof} should arise whether or not a domain wall forms after the collision.  One can see this from the analysis performed in Ref.~\cite{Aguirre:2007wm} in the pure vacuum, thin-wall limit.  In both cases, to conserve energy, a radiation shell will propagate into the post-collision region of the observation bubble.  This radiation shell will distort the geometry to the future of the collision event, leading to the advanced or delayed end of inflation in the post-collision spacetime from which one effectively obtains the profile Eq.~\ref{eq:prof}.    

\subsection{The Frame Shift} \label{subsec:frame}

The derivation of the temperature profile Eq.~\ref{eq:prof} is valid for scenarios in which the observer is born comoving with respect to the unperturbed portion of the last scattering surface.  In this case, the perturbation from the collision enters the past lightcone of the observer at late times.  However, it was shown in Ref.~\cite{Aguirre:2008wy} that $O(3,1)$-invariance can be spontaneously re-generated in the region to the future of a collision, thus ``just as many" observers can potentially form comoving with the \emph{perturbed} portion of the last scattering surface.  (Such observers were dubbed ``foreign-born" in \cite{Aguirre:2009ug}.)  As described below, these observers are typically those that would see a bubble with $\rho > 0$ (covering more than half of the sky).  By the arguments of the last section, most observers who witness any bubble collisions at all will see many with $\rho > 1$, and hence be of this type.

We argue, however, that the effects of $\rho > 1$ bubbles will be largely invisible, and that the effects of $\rho > 0$ bubbles will be degenerate with (and convertible into) those of $\rho < 0$ bubbles.
%Previous studies \cite{Chang:2008gj, Czech:2010rg,Kleban:2011pg} focused only on observers that formed outside the affected post-collision region (``native born" observers).  However, accounting for foreign-born observers alters the profile Eq.~\ref{eq:prof} since for this class of observers, the perturbation to the gravitational potential (which we assume to be adiabatic) was superhorizon at early times.  As a result, its contribution to the CMB temperature profile (which would be primarily dipole) will have been exactly cancelled in our approximation, as shown in Ref.~\cite{Erickcek:2008jp}.  This is because for these observers, one must go beyond the Sachs-Wolfe approximation and account for the contribution of the adiabatic superhorizon perturbation to their peculiar velocities, in which case the intrinsic CMB dipole will be precisely cancelled by the Doppler dipole.  This effect can be thought of as a ``frame shift" between observers born in the rest frame of the unperturbed portion of the LSS and those developing in the rest frame of the affected region.  As we discuss below, bubbles taking up large portions (or all) of the LSS will fall into this category, and since for $N_{\rm obs}\geq1$ we expect many accompanying bubbles with $\rho>1$, it is important to understand this effect.
To see this, we can adapt the arguments set forth in Ref.~\cite{Erickcek:2008jp} as follows:  in a $\Lambda$CDM universe, the total temperature anisotropy in a direction $\hat{\mathbf{n}}$ on the sky is given by a sum of the Sachs-Wolfe (including integrated effects) and Doppler anisotropies in conformal Newtonian gauge \begin{gather}\label{eq:sw_superhor} \left[\frac{\Delta T}{T}(\hat{\mathbf{n}})\right]_{SW+ISW}\sim \Phi(\tau_{ls},\hat{\mathbf{n}}\xi_{ls})+ \int_{\tau_{ls}}^{\tau_0}\frac{d \Phi}{d\tau}d\tau \\ \label{eq:dop_superhor}  \left[\frac{\Delta T}{T}(\hat{\mathbf{n}})\right]_{D}\sim\hat{\mathbf{n}}\cdot \left[\vec{v}(\tau_0,\vec{0})-\vec{v}(\tau_{ls},\hat{\mathbf{n}}\xi_{ls})\right] +\mathcal{O}(v^2) \end{gather} where the velocity $\vec{v}$ induced by a superhorizon perturbation is given by \begin{equation} \label{eq:vel_superhor} \vec{v}(\tau,\vec{x})\sim \nabla \Phi + \frac{d}{d\ln a}\nabla \Phi \end{equation} (note that the we have neglected the curvature of the last scattering surface, a good approximation for $\xi_{ls}\ll1$).  Since the above expressions for the temperature difference are linear in $\Phi$ and its derivatives, to lowest order in velocity and considering only the linear piece of $\Phi(\vec{x})$, the various contributions for a gravitational potential of the form $\Phi=\Phi_1+\Phi_2$ are simply the sum of the individual contributions: \begin{equation} \label{eq:pert_sum}  \frac{\Delta T}{T}(\hat{\mathbf{n}}) =  \frac{\Delta T}{T}(\Phi_1)+ \frac{\Delta T}{T}(\Phi_2) .\end{equation}  

Consider a particular collision intersecting the observer's LSS with angular scale $\psi$.  Neglecting the curvature of the last scattering surface, one can imagine the collision inducing a planar perturbation \begin{equation} \label{eq:pert} \Phi(x)=\bar{\alpha}(x-x_{int})\Theta(x-x_{int}) \end{equation} at $\tau_{ls}$, where both nucleation centers lie along the $x$-axis, and $x_{int}$ denotes the intersection of the future light cone of the collision with the LSS.  
%The observer's LSS is determined by the intersection of his or her past light cone with the $H_3$ at $\tau=\tau_{ls}$.  
One can rewrite Eq.~\ref{eq:pert} as a superposition of two modes $\Phi_1(x)$, $\Phi_2(x)$ as \begin{equation} \Phi(x)=\bar{\alpha}(x-x_{int}) + \bar{\alpha}(x_{int}-x)\Theta(x_{int}-x) \equiv \Phi_1(x)+\Phi_2(x) .\end{equation}  Here $\Phi_1(x)$ is a superhorizon mode while $\Phi_2(x)$ is equivalent to that of a collision centered around the antipodal point of the original collision on the celestial sphere with angular scale $\psi'=2\pi-\psi$ and strength $\bar{\alpha}$.  

A foreign born observer emerging from the last scattering surface at $x>x_{int}$ that is initially comoving with the perturbed LSS only sees the superhorizon mode $\Phi_1$ at early times, with $\Phi_2$ only entering his or her past light cone at late times.  As a result, the contribution of $\Phi_1$ to the observed CMB temperature spectrum will vanish as in Ref.~\cite{Erickcek:2008jp} and using Eq.~\ref{eq:pert_sum}, the resulting CMB temperature fluctuation will be \begin{equation} \label{eq:frame_shift} \frac{\Delta T}{T}= \frac{\Delta T}{T}(\Phi_2) .\end{equation}  The effect of a collision on the CMB temperature profile for a foreign-born observer from a collision centered at $(\theta_0,\phi_0)$ with angular scale $\psi$ is therefore equivalent to the profile seen by a native-born observer for a collision centered at the antipodal point $(\pi-\theta_0,2\pi-\phi_0)$ with scale $\psi'=2\pi-\psi$; the observed perturbation is shifted by going between the native and foreign born observers' reference frames. 

%For a foreign-born observer at early times, the collision results in a superhorizon perturbation, while at later times the unperturbed portion of the LSS begins to enter his or her past light cone.   the collision affects more than half of the last scattering surface ($\psi>\pi$) and the perturbation to the gravitational potential can be written as the sum of a superhorizon perturbation and a contribution from a collision centered at the antipodal point with angular size $\psi'=2\pi-\psi$  The superhorizon contribution is governed by Eqs.~(\ref{eq:sw_superhor},\ref{eq:dop_superhor},\ref{eq:vel_superhor}) and as shown in Ref.~\cite{Erickcek:2008jp} does not contribute to the CMB temperature profile to lowest order: $\Delta T/T(\Phi_1)=0$; thus \begin{equation} \label{eq:frame_shift} \frac{\Delta T}{T}= \frac{\Delta T}{T}(\Phi_2) \end{equation}   The effect of a collision on the CMB temperature profile for a foreign-born observer from a collision centered at $(\theta_0,\phi_0)$ with angular radius $\psi_0/2$ is equivalent to the profile seen by a native-born observer for a collision centered at the antipodal point $(\pi-\theta_0,2\pi-\phi_0)$ and angular radius $\psi'/2=\pi-\psi_0/2$.

The above argument applies to observers whose past worldlines intersect the perturbed portion of the last scattering surface for a given collision.  This will clearly be the case for collisions encompassing the observer's entire LSS ($\rho>1$),  whose perturbations consequently vanish in our approximation.  This allows us to neglect the potentially large number of full-sky collisions accompanying $N_{\rm obs}\ge 1$ in calculating the CMB temperature power spectrum.\footnote{It is possible that there are other observable effects associated with collisions on these scales.  We do not consider this possibility here and in the remainder of this work simply neglect bubbles with scales $\rho>1$, cautioning the reader that the compatibility of some large-$N_{\rm obs}$ scenarios with observation may be spoiled by the inclusion of such effects.}  

%%%%%%%%%%%%%%%%%%%%%%%%%%%%%%%%%%%%%%%%%%%%%%%%%%%
\begin{figure}[!t]
\mbox{\includegraphics[width=0.45\textwidth,clip]{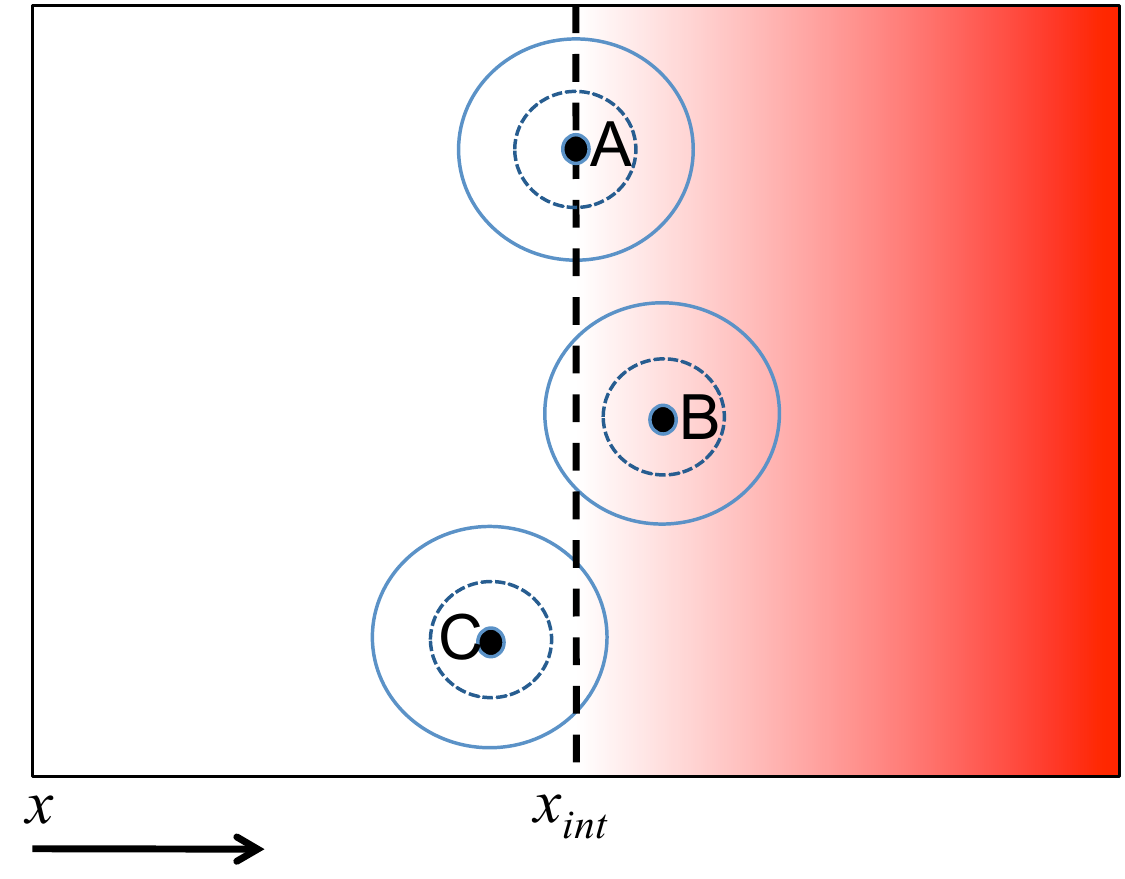}}
\caption{\label{fig:frame_shift}\small The universe with one bubble collision at decoupling.  We show three different observers who see a collision with profile Eq.~\ref{eq:pert} taking up (A) half of the sky, (B) more than half of the sky, and (C) less than half of the sky.  Their present-day last scattering surfaces are shown by the blue circles, corresponding to the intersection of their past lightcones with the decoupling time slice.   Unless the observers have acquired a large peculiar velocity along the $x$ direction, the last scattering surfaces at some earlier time are approximately given by the dashed blue lines for each observer.  In this case, observer B is foreign born and sees a frame shifted collision while C does not.  Observer A is a boundary case and could have emerged from either portion of the LSS.  In our approximation, we take all observers born with $x>x_{int}$ to observe a frame shifted collision}
\end{figure}
%%%%%%%%%%%%%%%%%%%%%%%%%%%%%%%%%%%%%%%%%%%%%%%%%%%

This ``frame shift" will also tend to apply to collisions with $\rho<1$ that cover more than half of the observer's sky.  For a native born, comoving observer originating from $x\approx x_{int}$, the collision will nearly bisect the LSS.  This situation is depicted by observer A in Fig.~\ref{fig:frame_shift}.  As long as a foreign-born observer's worldline does not intersect those of native comoving $x\approx x_{int}$ observers (which should typically not occur, since late time peculiar velocities will be dominated by the effects of processes uncorrelated with the collision itself), the foreign born observer will see a greater-than-half sky collision (observer B in Fig.~\ref{fig:frame_shift}).  Conversely, native-born observers will only see a collision with $\rho>0$ if they also acquire a significant peculiar velocity in the direction of the collision (c.f. observer C in Fig.~\ref{fig:frame_shift}).  The farther away from $x_{int}$ the observer is born, the larger the peculiar velocity will need to be for him or her to observe a greater than half-sky bubble that is not frame shifted.  Thus we expect that most $\rho>0$ collisions will result in frame-shifted profiles, appearing to take up \emph{less} than half of the observer's sky, while most with $\rho<0$ will be unaffected.  

For illustrative purposes we will work under the assumption that all greater-than-half sky bubbles correspond to frame-shifted collisions.  We implement this in the bubble profile by mapping all such collisions to less-than-half sky events centered around the antipodal point of the original nucleation center on the sky.  Relaxing this assumption does not significantly affect our conclusions, since this prescription primarily effects the dipole, reducing it compared to the case where the frame shift is not taken into account. 

\section{Multiple Collisions: Generalizing To $N_{\rm obs}>1$}\label{sec:mult_prof}

\subsection{A Multiple Collision Scenario As The Sum Of Its Parts} \label{subsec:intervening}

We wish to generalize the results in Secs.~\ref{sec:distributions}-\ref{sec:profile} to the case of $N_{\rm obs}>1$.  Colliding bubbles that do not overlap in their influence prior to the last-scattering surface can simply be treated as independent.  Events where bubbles do overlap  prior to this time can divided into ``pre-collisions" occurring before either bubble intersects the observation bubble, and overlaps occurring within the observation bubble, prior to last-scattering.  We will first show that the first category are uncommon and hence generally unimportant, then discuss the second category.

%In particular, we must address whether or not the effects of multiple collisions on the observation bubble can be thought of as a collection of single collision events described by the analysis in the previous sections, or whether new features associated with multiple bubbles significantly alter the picture.  We assume that the combined effects of all collisions allow for a sufficient period of inflation everywhere within the bubble, so that any post-collision effects can be treated perturbatively after a few e-folds.  We expect this to hold in most cases of interest, since the collisions should be compatible with our observed cosmology.
 
%In scenarios with $N_{\rm obs}>1$, any given collision can be analyzed with the methods described above for a single collision, at least until the affected post-collision regions from two or more events come into causal contact with each other.  However, some bubbles may collide with each other \emph{before} hitting the observation bubble wall.  If this were to occur, the observer would see several dbefisturbances entering his or her bubble from these ``pre-collisions" along with the effects of bubbles directly impacting the observation bubble wall.  The resulting CMB profile would have to account for these two types of events.

\begin{figure*}[!t]
\mbox{\hspace*{-1.2cm}\includegraphics[width=0.3\textwidth,clip]{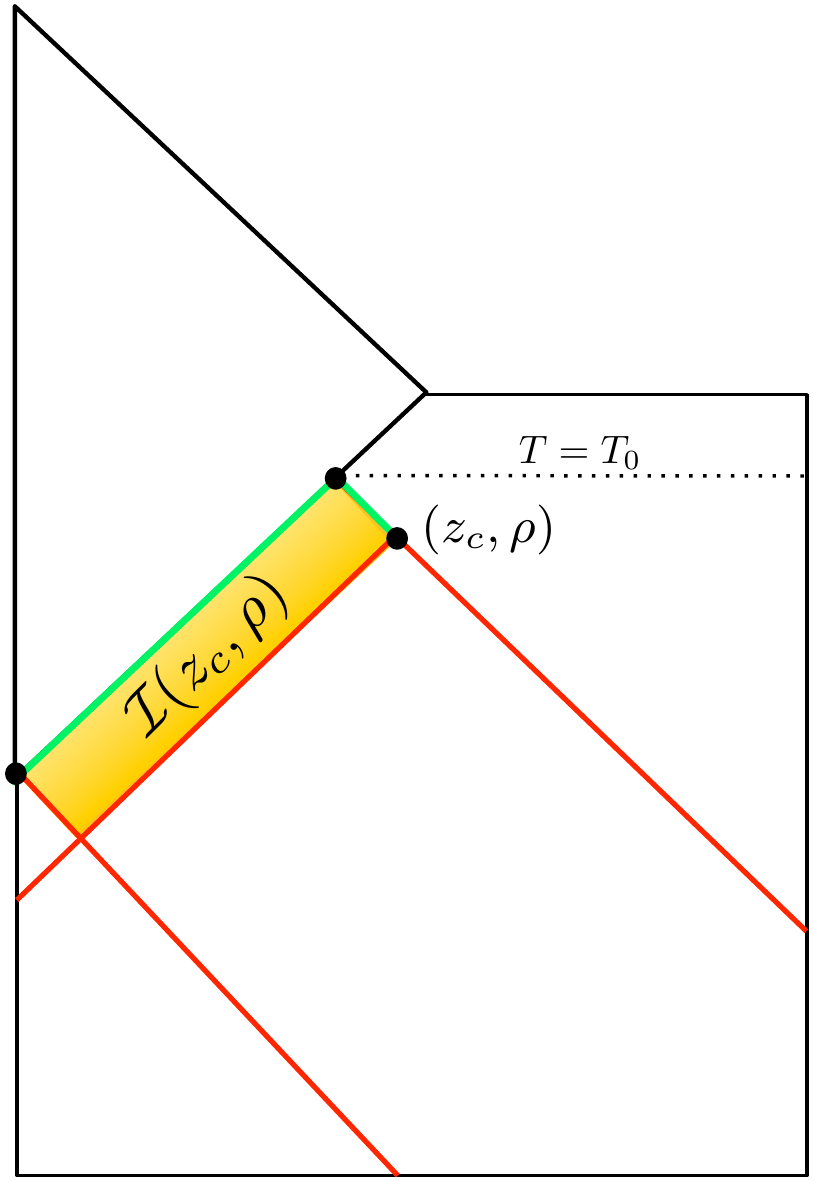}}
\caption{\label{fig:int_conformal}\small A conformal slice of the 4-volume available to nucleate an intervening bubble for a collision bubble nucleated at $(z_c,\rho)$ (with $\theta=\phi=0$).  The region $\mathcal{I}(z_c,\rho)$ is given by the 4-volume within the past lightcone from the point on the observation bubble wall with $T=T_0$ (in green), excluding the the 4-volume within the past lightcones from $(z_c,\rho)$ and the origin shown in red.}
\end{figure*}

%However, for typical eternal inflation scenarios and $H_{I}/H_{F}\ll1$, an observer will see the effects of very few pre-collisions.  To see why this is the case, 
Consider the disturbance from a pre-collision impacting the observation bubble wall at $T_0$ in the observation frame.  We can model this disturbance as a shell of radiation emanating from the pre-collision event.  Observing this radiation shell implies that one of the $N_{\rm obs}$ colliding bubbles, which would have impacted the observation bubble at $T_0$, was intercepted by an intervening bubble before it could do so.  For a given bubble with nucleation center\footnote{As discussed in Appendix~\ref{sec:Ap_1}, we are free to choose $\theta_c=\phi_c=0$ without loss of generality.} $(z_c,\rho)$ colliding with the observation bubble wall at $T_0$, we can ask how many intervening bubbles of a particular type it expects to encounter along the way to the observer's bubble.  We denote this quantity as $N_{\rm int}( z_c,\rho)$; for a fixed $\lambda$, it is given by \begin{equation}\label{eq:N_int} N_{\rm int} (z_c, \rho)=\lambda \hspace{1mm} \mathcal{I}(z_c,\rho) \end{equation} where a conformal slice of the 4-volume $\mathcal{I}(z_c,\rho)$ is depicted by the region shaded yellow in Fig.~\ref{fig:int_conformal}.  Note that $N_{\rm int}$ includes bubbles with $\rho>1$ in its count, and that these bubble need not nucleate at the same $(\theta,\phi)$ values as the collision bubble.

%%%%%%%%%%%%%%%%%%%%%%%%%%%%%%%%%%%%%%%%%%%%%%%%%%%
%%%%%%%%%%%%%%%%%%%%%%%%%%%%%%%%%%%%%%%%%%%%%%%%%%%

%%%%%%%%%%%%%%%%%%%%%%%%%%%%%%%%%%%%%%%%%%%%%%%%%%%
\begin{figure}[!t]
\mbox{\includegraphics[width=0.45\textwidth,clip]{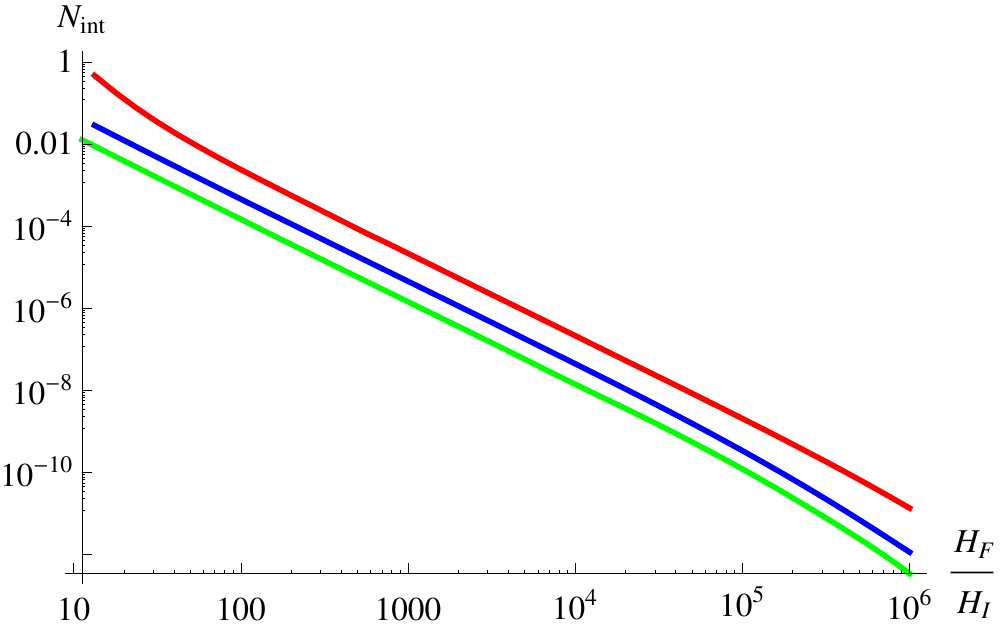}}
\caption{\label{fig:lambda_min}\small The expected number of intervening bubbles as a function of $H_F/H_I$ for one collision bubble ($N_{\rm obs}=1$) nucleating from $z_c=0.5,1,5$ (green, blue, red) and $\rho=1$ in the false vacuum and with $\xi_{ls}=.05$.  Varying $\rho$ in the interval $\left[-1,1\right]$ has little effect on the curves, as expected from the discussion surrounding Fig.~\ref{fig:conformal_2}.  Increasing $N_{\rm obs}$ will shift these curves upward, however for $H_F/H_I\gg1$ we still expect $N_{\rm int}\ll1$.}
\end{figure}
%%%%%%%%%%%%%%%%%%%%%%%%%%%%%%%%%%%%%%%%%%%%%%%%%%%

The 4-volume $\mathcal{I}(z_c,\rho)$ corresponds to the portion of the past light cone from the would-be intersection point of the collision bubble with the observation bubble wall, $(\eta_0,T_0)$, not also within the past light cones of $(z_c, \rho)$ and the origin.  Intervening bubbles nucleated outside of the past light cone from $(\eta_0,T_0)$ cannot produce radiation from the pre-collision observed at $T_0$, while requiring that the collision and observation bubbles were not born within the intervening bubble eliminates the 4-volume in the past light cones from $(z_c,\rho)$ and the origin, respectively.  Defined this way, $\mathcal{I}(z_c,\rho)$ will not include pre-collisions whose radiation enters the observation bubble farther up the $H_2$ collision surface (i.e. $T>T_0$), but these events will be counted in $N_{\rm int}$ for the corresponding \emph{intervening} bubble.  Thus, for a given $N_{\rm obs}$, we can look at $N_{\rm int}$ for each observable bubble and if this quantity is small in all cases, we can safely neglect the effects of all pre-collisions on the effective temperature profile.

Using the 4-volume element in Eq.~\ref{eq:4vol}, we can estimate $N_{\rm int}$ for a given collision bubble nucleated at $(z_c,\rho)$.  This calculation is detailed in Appendix~\ref{sec:Ap_1}.  The results are illustrated in Fig.~\ref{fig:lambda_min} where $\lambda$ is determined by requiring $N_{\rm obs}=1$.  For large $H_F/H_I$, the curves of constant-$\rho$ for a given $z_c$ are compressed as in Fig.~\ref{fig:conformal_2} and so the curves in Fig.~\ref{fig:lambda_min} are nearly identical for all $\rho$ in the interval $[-1,1]$.  

%The radiation shell from a pre-collision travels within the future light cone of the intervening bubble nucleation point and so 
%To do so we must first verify that any of the individual $N_{\rm obs}$ collision events is qualitatively similar to the case where $N_{\rm obs}=1$.  In particular, when many bubbles impact the observation bubble wall one may worry that several of these incoming bubbles will collide with each other first.  If this is the case, then what impacts the observation bubble is not the colliding bubble wall but instead the radiation shell released by the ``pre-collision" and possibly the domain wall formed between the two pre-colliding bubbles if they are associated with different vacua.  As a result of the underlying symmetries, these effects may result in similar observational signatures to those associated with a single bubble collision, however they are qualitatively different events and would require greater care when computing expectation values as in Sec.~\ref{} since they introduce an extra parameter into the perturbation profile, namely the dS-invariant distance between the colliding bubbles as well as between the pre-collision and the observation bubble.

From Fig.~\ref{fig:lambda_min} we see that most bubbles will indeed not collide with an intervening bubble.  This is because although many bubbles can nucleate as $T\rightarrow \pi/2$, very few of them will have a chance to intersect the collision bubble wall before it impacts the observer's bubble.  The past light cone from the intersection point of the collision bubble with the observation bubble wall contains very little 4-volume near the future infinity of dS, only opening up to enclose more 4-volume for smaller $T$.  As a result, regions for which $dV_4$ starts to diverge do not contribute very significantly to $\mathcal{I}$.  Note also that smaller values of $z_c$ correspond to smaller values of $N_{\rm int}$.  This is because for large $H_F/H_I$ and the small $\xi_{ls}$ we consider, the collision bubble nucleation points corresponding to small $z_c$ are very close to the observation bubble wall, resulting in a smaller 4-volume $\mathcal{I}(z_c,\rho)$ and hence smaller $N_{\rm int}$.  For larger $z_c$, this cancellation is not as severe and $N_{\rm int}$ can be close to 1 for small $H_F/H_I$, however this will necessitate a large $\lambda$ which is in tension with the requirement for eternal inflation.  Also, increasing $N_{\rm obs}$ will shift the curves upward, however for cases of interest with $H_F/H_I\gg1$, $\xi_{ls}\ll1$, we will still expect $N_{\rm int}\ll1$ unless $N_{\rm obs}$ is very large as well.  Thus, in what follows we can safely neglect any interactions between the bubbles before they impact the observation bubble wall.

Although given our parameters of interest, bubbles only rarely ``pre-collide", they should commonly overlap \emph{within} the observation bubble if $N_{\rm obs} \gg 1$, especially considering $\rho > 1 $ (``all sky") collisions.  If these overlaps occur after the effects of the collisions have entered the perturbative regime, the various perturbations will simply superpose.  This will not necessarily be the case for large $N_{\rm obs}$, however.  Consider the 2-sphere defined by the intersection of the observer's past light cone with the observation bubble wall.  Given a uniform distribution of collision bubble nucleation centers, the points of intersection of the collision bubbles with the $S_2$ will be distributed with average distance $\sim \pi R_0 /N_{\rm obs}$ separating any two of these points, where $R_0$ is the observation bubble radius at the time of intersection.  Since $R_0\sim H_I^{-1}$, one would expect the future light cones from these points to intersect early on for large $N_{\rm obs}$, and so in principle the effects of the various collisions may interact with each other before enough inflation has occurred to dilute the disturbances to the point of being perturbative.  

To study these interactions would require details of the microphysics governing the collisions, as in the numerical study of Ref.~\cite{Johnson:2011wt}.  Such considerations are beyond the scope of this paper, and so in what follows we neglect these interactions: in treating the effects of the collisions inside the observation bubble, we approximate the various disturbances as simply superposing on one another.  The resulting effects on the CMB temperature will then be described by the sum of Sachs-Wolfe contributions from the individual collisions (including the frame shift).  We intend to test this approximation in future numerical studies, but for the time being we content ourselves with this treatment, as we expect it to characterize (at least approximately) the effects of overlapping collisions on the CMB.

With the above assumptions, a multiple collision scenario can indeed be analyzed as a collection of single collision events.  Accounting for the frame shift (and thus neglecting bubbles with $\rho>1$), the resulting CMB temperature profile for a given multiple collision scenario can be written as \begin{equation} \label{eq:multi_prof} \begin{aligned} T(\hat{\mathbf{n}}) =&T^{\prime}_0\left(1+\Delta(\hat{\mathbf{n}}) \right) \\  & \times \Big[1+ \sum_{N_1}\alpha_i \xi_{ls}(\mu_{i}-\cos\frac{\psi_i}{2})\Theta(\mu_{i}-\cos\frac{\psi_i}{2})   \\ &+ \sum_{N_2}\alpha_i \xi_{ls}(-\mu_{i}+\cos\frac{\psi_i}{2})\Theta(-\mu_{i}+\cos\frac{\psi_i}{2}) \Big]  \end{aligned} \end{equation} where we have defined the angular distance for a given collision as \begin{equation} \mu_{i}=\sin\theta \sin\theta_{0i} \left(\cos\phi \cos \phi_{0i} + \sin\phi \sin\phi_{0i}\right) + \cos\theta \cos\theta_{0i} \end{equation} and  where $(\theta_{0i}, \phi_{0i}$) is the angular position of the nucleation center of the $i$th bubble on the observer's sky, $N_1$ ($N_2$) is the number of bubbles affecting less than (more than) half the sky, $\alpha_i$ is the inherent brightness of a given collision, $T_0'$ is the average temperature neglecting any perturbations, and we have included the inflationary density perturbations $\Delta(\hat{\mathbf{n}})$ which we approximate to be unaffected by the collisions.  Note that in order to treat effects of the collisions on the CMB temperature perturbatively as we have done, we should have $N_{\rm obs} \alpha_i \xi_{ls} \ll 1$ for all collisions.  With a few additional assumptions we can use Eq.~\ref{eq:multi_prof} to extract some general features of the effects of multiple collision scenarios on the CMB. 

\subsection{Additional Assumptions}\label{sec:assumptions}

Several more useful observations and simplifications can applied to the picture above: 
\begin{itemize} 
\item We neglect the possibility of ($\rho < 1$) bubbles forming within other (late time) bubbles. 
% For a given point $(\eta,T)$ in the false vacuum, the 4-volume out of which a bubble could have nucleated such that $(\eta,T)$ is contained in this bubble is just the 4-volume in the past light cone from $(\eta,T)$.  
For small $\lambda$ this is reasonable, for similar reasons as the neglect of pre-collisions: the overlap between the past of an observer's last-scattering surface and the future of a given bubble collision is generally of order $H_F^{-4}$, so the number of nucleations from that region, which would consitute bubbles-in-bubbles, is small unless $\lambda > H_F^{-4}$.
%However, by the discussion surrounding Fig.~\ref{fig:lambda_min}, unless the corresponding nucleation rate is large, such events will be rare and so we can safely neglect bubbles-within-bubbles in our scenario.  
Arguments along these lines have been realized in the past (see e.g. Refs.~\cite{Freivogel:2009it, Aguirre:2007an}). 
\item We assume that only one decay channel is relevant for nucleating colliding bubbles from the parent vacuum.  For a long-lived vacuum this does not seem unreasonable: since $\lambda=Ae^{-S_E}$, where $S_E$ is the Euclidean action of the CDL instanton and $A$ is a prefactor containing quantum corrections, field trajectories resulting in small differences in $S_E$ will tend to engender rather large differences in $\lambda$ so that one dominates.  For a recent discussion along this line of thought, see Ref.~\cite{Guth:2012ww}.  This assumption allows us to consider only one type of collision bubble relevant for the multiple collision scenario. 
\item We imagine that the potential $V(\varphi)$ in the neighborhood of $\varphi$ after any of the collisions is relatively featureless, i.e. $V(\varphi+\delta \varphi_i)\approx V(\varphi+\delta\varphi_j)$ and likewise for $V'$.  Since we require a period of slow roll inflation to the future of all collisions, which implies a flat potential near the post collision field values, this assumption is typically satisfied \textit{a posteriori}.  We note that it is difficult to see how this assumption, or even that of a multiple collision scenario, could be satisfied for small-field inflationary models (i.e. models where the width of the inflationary trajectory $\Delta \varphi<M_{pl}$) without very small values of $\delta \varphi$, since even small perturbations will tend to cause the inflaton to overshoot the inflationary region of $V(\varphi)$ \cite{Johnson:2011wt}.\end{itemize} 

The main consequence of the above assumptions for our purposes is that the brightness parameters $\alpha_i$ only depend on $z_c$ and $\rho$, by the following reasoning.  For a fixed potential with colliding bubbles of only one type, the strength $\bar{\alpha_i}$ of the perturbation at $\tau_{rh}$ depends on the kinematics through $\delta\varphi$.  In the collision frame, $\delta \varphi$ is determined only by the parameter $z_c$, while the boost back into the observation frame generally depends on $\rho$.   However, since we are most interested in cases where $H_F/H_I\gg1$, collision bubbles with $\rho$ in the interval $[-1,1]$ are all mapped to a very small range of $\eta$, $T$ for a given $z_c$, and hence all receive virtually the same boost.  Also, $\alpha$ is related to $\delta \varphi$ by factors which depend on the background cosmology as well as $V(\varphi)$, $V'(\varphi)$, but by the third assumption above, this factor should be similar for all collisions. Consequently, $\alpha\simeq\alpha(z_c)$ and the brightness parameters in Eq.~\ref{eq:multi_prof} in our approximation depend only on the kinematic variable $z_c$ and the underlying potential, up to some overall normalization.  We will exploit this fact in taking observer expectation values for the temperature spectrum in the following section.

\section{Effects On CMB Temperature Power Spectrum} \label{sec:power_spect}

Given that, in many cases of interest, the effects of multiple bubble collisions may be approximated by superposition, it is worth considering the observational consequences of such scenarios by generalizing previous work to the case of many bubbles. Armed with the temperature profile Eq.~\ref{eq:multi_prof}, the remainder of the paper constitutes a first look at multiple collision signatures, by computing their effects on the CMB temperature power spectrum.

Our setup and framework follows that suggested in Refs.~\cite{Gordon:2005ai, Chang:2008gj}, with the appropriate generalizations to multiple collisions. Denoting the sky-averaged observed temperature as $T_0$, we expand the observed temperature fluctuations for a given set of collisions in terms of spherical harmonics 
\begin{equation}\label{eq:contrast} 
\frac{\delta T(\hat{\mathbf{n}})}{T_0}=\frac{T-T_0}{T_0}=\sum_{l,m}t_{lm}Y_{lm}(\hat{\mathbf{n}}) .
\end{equation}  
We define
 \begin{equation} 
 A(\hat{\mathbf{n}})\equiv f(1+w(\hat{\mathbf{n}}))-1\end{equation} \vspace{-8mm} 
 \begin{equation}  B(\hat{\mathbf{n}})\equiv \Delta(\hat{\mathbf{n}}) \end{equation} \vspace{-8mm} \begin{equation} \begin{aligned}w(\hat{\mathbf{n}})& \equiv \sum_iw^i(\hat{\mathbf{n}})=\sum_{N_1}\alpha_i \xi_{ls}(\mu_{i}-\cos\frac{\psi_i}{2})\Theta(\mu_{i}-\cos\frac{\psi_i}{2}) \\ & +\sum_{N_2}\alpha_i \xi_{ls}(-\mu_{i}+\cos\frac{\psi_i}{2})\Theta(-\mu_{i}+\cos\frac{\psi_i}{2}) \end{aligned} \end{equation} where $f=T_0^{\prime}/T_0$, and $\Delta(\hat{\mathbf{n}})$ is the standard inflationary temperature fluctuation in direction $\hat{\mathbf{n}}$. (We assume $T_0=2.725$ K throughout our calculations.)

 Using these definitions we can recast Eq.~\ref{eq:contrast} as \begin{equation}  \frac{\delta T(\hat{\mathbf{n}})}{T_0}=  A(\hat{\mathbf{n}}) + f\left(1+w(\hat{\mathbf{n}})\right)B(\hat{\mathbf{n}}) .\end{equation}   Like the temperature contrast of Eq.~\ref{eq:contrast},  we can expand the functions $w$, $A$, $B$ in spherical harmonics with corresponding coefficients $w_{lm}$, $a_{lm}$, $b_{lm}$.  Then, using the properties of products of spherical harmonics, the coefficients $t_{lm}$ are given by \begin{equation} \label{eq:tlm} t_{lm}= a_{lm}+f b_{lm}+ f \sum_{l_1m_1}\sum_{l_2m_2}w_{l_1m_1}b_{l_2m_2}R_{lm}^{l_1m_1l_2m_2}\end{equation} where the $R$ terms are Gaunt coefficients, given in terms of Wigner $3-j$ symbols by \begin{equation} \begin{aligned} R_{lm}^{l_1m_1l_2m_2}=&(-1)^m\sqrt{\frac{(2l_1+1)(2l_2+1)(2l+1)}{4\pi}}  \\ &\times \left( \begin{array}{ccc} l_1 &  l_2 & l \\ 0 & 0 & 0 \end{array} \right) \left( \begin{array}{ccc} l_1 &  l_2 & l \\ m_1 & m_2 & -m \end{array}\right) \end{aligned} .\end{equation}    The conventions we use here are those found in Ref.~\cite{book}, to which we refer the reader for additional details.  Note that the isotropy in the corresponding expressions in Ref.~\cite{Chang:2008gj} is broken as a result of including more than one collision.  

%[[edited to here.]]
\subsection{The Two-Point Function}

We are interested in the expected angular temperature power spectrum for a CMB sky affected by many bubbles.  To compute this quantity, one must evaluate the two-point function $\left<t_{lm}t^*_{lm}\right>$, using Eq.~\ref{eq:tlm} to average over both the ensemble of density fluctuations from inflation (hereafter ``ensemble averages"), as well as over the distributions governing the various collision parameters.
 
\subsubsection{Ensemble Averages}

First let us perform the ensemble averages.  Consider a given set of collisions $\mathcal{B}$ arising from a fixed potential and $N_{\rm obs}$.  Following previous work, we approximate the inflationary density perturbations as being unaffected by the collisions.  The only terms affected by the ensemble average for a given set of collisions are those containing $b_{lm}$ in Eq.~\ref{eq:tlm}.  Since the density perturbations are assumed to be Gaussian with zero mean, the one point functions $\left<b_{lm}\right>^{\mathcal{B}}$ vanish for any given set of collisions and only terms with two-point functions $ \left<b_{lm}b^*_{lm}\right>^{\mathcal{B}}$ will survive (here the brackets $\left<\right>^{\mathcal{B}}$ denote ensemble averages with $\mathcal{B}$ fixed).  If the density perturbations are unaffected by the collisions, the corresponding two-point function is given by \begin{equation} \left<b_{lm}b_{lm}^*\right>^{\mathcal{B}}=C_{l}^{bb} \end{equation} where $C_{l}^{bb}$ is the correlation function in the absence of any collisions (note that $\left<b_{l_1m_1}b_{l_2m_2}^*\right>^{\mathcal{B}}=0$ for $l_1\neq l_2$).  We obtain the $C_l^{bb}$ from {\tt CMBEASY} \cite{Doran:2003sy} using concordance WMAP7 values \cite{WMAP} for the relevant cosmological parameters.  Performing the ensemble average for this set of collisions yields \begin{equation} \label{eq:t_lm} \begin{aligned} \left<t_{lm}t^*_{lm}\right>^{\mathcal{B}} =&  a_{lm}a^*_{lm}+ f^2\Big( C_{l}^{bb}+ \\ &\sum_{l_1m_1}w_{l_1m_1}R_{lm}^{l_1m_1lm}C_{l}^{bb} + cc. + \\ & \sum_{l_{i}m_{i}}w_{l_1m_1}w^*_{l_3m_3}R_{lm}^{l_3m_3l_2m_2}R_{lm}^{l_1m_1l_2m_2}C_{l_2}^{bb}\Big) \end{aligned} \end{equation}  where the sum in the third line is over $i=1,2,3$.  From Eq.~\ref{eq:t_lm} we can calculate the CMB angular temperature power spectrum $C^{\mathcal{B}}_l$ for a set of collisions $\mathcal{B}$ via \begin{equation} \label{eq:cl_def} C^{\mathcal{B}}_l=\frac{1}{2l+1}\sum_m\left<t_{lm}t_{lm}^*\right>^{\mathcal{B}} .\end{equation}   

\subsubsection{Averages Over Collision Scenarios}

To understand the generic features of a multiple collision scenario, we wish to average Eqs.~\ref{eq:t_lm}, \ref{eq:cl_def} over the set of possible collision scenarios, $\left\{\mathcal{B}(N_{\rm obs})\right\}$ with the underlying potential and $N_{\rm obs}$ fixed.  This yields \begin{equation} \label{eq:t_lm_bub} \begin{aligned} \left<t_{lm}t_{lm}^*\right> =&  \left<a_{lm}a^*_{lm}\right>+ \left<f^2\right>\Big( C_l^{bb}+ \\ &\sum_{l_1m_1}\left<w_{l_1m_1}\right>R_{lm}^{l_1m_1lm}C_l^{bb} + cc. + \\ & \sum_{l_{i}m_{i}}\left<w_{l_1m_1}w_{l_3m_3}^*\right>R_{lm}^{l_3m_3l_2m_2}R_{lm}^{l_1m_1l_2m_2}C_{l_2}^{bb}\Big) \end{aligned} \end{equation} where the brackets now denote averages over $\left\{\mathcal{B}(N_{\rm obs})\right\}$ for a given potential and $N_{\rm obs}$.  From this, one can compute the expected CMB angular temperature power spectrum via \begin{equation} \label{eq:cl_bub} \left<C_l\right>=\frac{1}{2l+1}\sum_m\left<t_{lm}t_{lm}^*\right> .\end{equation}

\begin{figure*}[!t]
\mbox{\includegraphics[width=0.45\textwidth,clip]{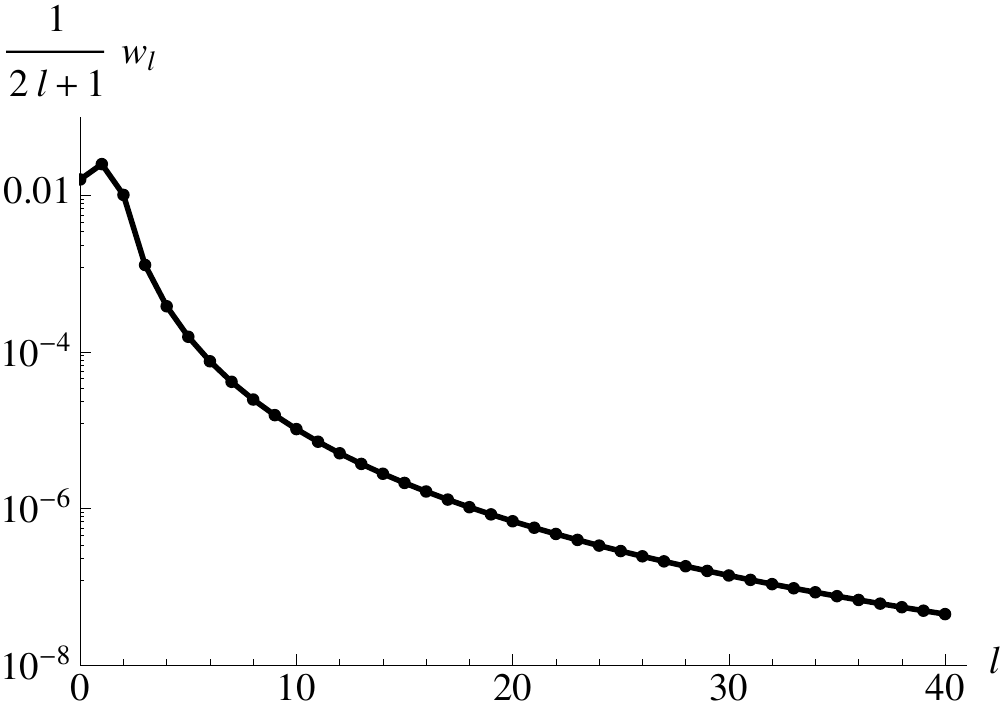}\qquad \includegraphics[width=0.45\textwidth,clip]{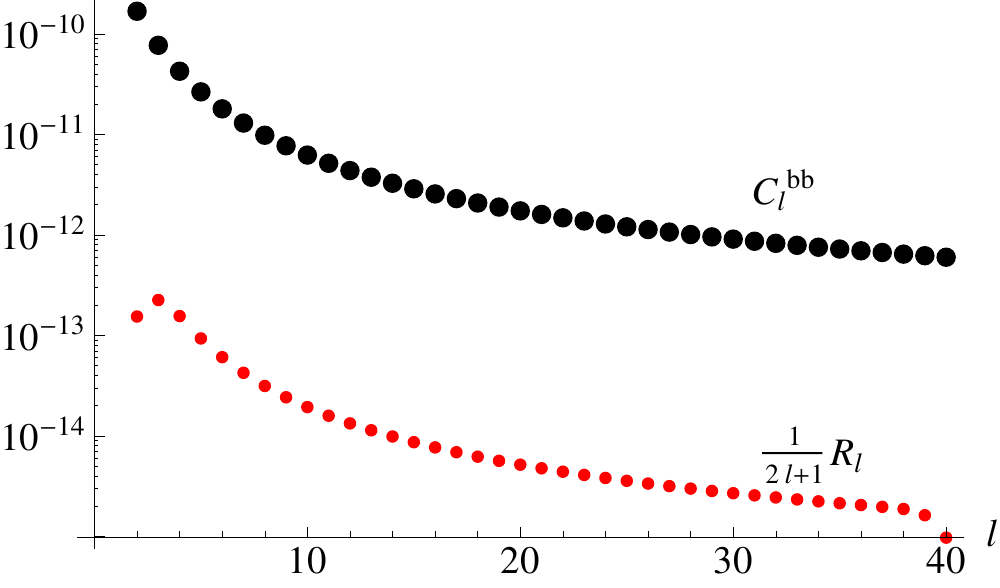}}
\caption{\label{fig:w_l}\small Left: The dimensionless quantity $w_l$ which enters into our final expression for $\left<C_l\right>$, plotted for $0\leq l\leq 40$.  The $l=0,1$ values are reduced relative to the case not accounting for the frame shift.  $w_l$ encodes the shape of the spectrum resulting for the isotropic and scale-invariant distribution of $\rho$; its contribution to $\left<C_l\right>$ is multiplied by $\pi^2 N_{\rm obs} \alpha_2 \xi_{ls}^2$.  Right: The red points denote the quantity $R_l$ from Eq.~\ref{eq:Rl} plotted for $2\leq l \leq 40$. $R_l$ also enters into Eq.~\ref{eq:spectrum} carrying a factor of $\pi^2 N_{\rm obs} \alpha_2 \xi_{ls}^2$.  Note that the dips for $l=2,40$ are a result of excluding $l_2<2$ and $l_2>40$ in the sum in Eq.~\ref{eq:Rl}.  The theoretical values $C_l^{bb}$ obtained for a concordance $\Lambda$CDM cosmology from {\tt CMBEASY} are shown by the large black points.}
\end{figure*}
%%%%%%%%%%%%%%%%%%%%%%%%%%%%%%%%%%%%%%%%%%%%%%%%%%%

\subsection{Performing The Averages}\label{subsec:averaging}

%%%%%%%%%%%%%%%%%%%%%%%%%%%%%%%%%%%%%%%%%%%%%%%%%%%

%\begin{figure*}[!t]
%\centering
%\subfigure[\small The quantity $w_l$ which enters into our final expression for $\left<C_l\right>$, plotted for $0\leq l\leq 40$.  The $l=0,1$ values are reduced relative to the case not accounting for the frame shift.  $w_l$ encodes the shape of the spectrum resulting for the isotropic and scale-invariant distribution of $\rho$; its contribution to $\left<C_l\right>$ is multiplied by $\pi^2 N_{\rm obs} \alpha_2 \xi_{ls}^2$.]{
%\includegraphics[scale=.8]{wl_plotting.pdf}
%\label{fig:w_l}
%}
%\subfigure[\small The red points denote the quantity $R_l+R_l^0(N_{\rm obs}$ for $N=100, 10^6$ from Eq.~\ref{} plotted for $2\leq l \leq 40$. The quantities shown enter into Eq.~\ref{eq:spectum} multiplied by a factor of The theoretical values $C_l^{bb}$ obtained for ]{
%\includegraphics[scale=.8]{Rl_Cbb_fig.pdf}
%\label{fig:Rl_Cbb}
%}
%\end{figure*}
%%%%%%%%%%%%%%%%%%%%%%%%%%%%%%%%%%%%%%%%%%%%%%%%%%%

%Consider now an observer who has $N_{\rm obs}$ collisions intersecting the visible portion of the LSS.  What is the expected signal in the CMB temperature power spectrum, $\left<C_l\right>$?  To answer this question, one must average over $\left\{\mathcal{B}(N_{\rm obs})\right\}$, or, equivalently, over the parameters $z_c$, $\psi$, $\theta$, $\phi$ for a given potential.
To perform the averages over the various $\mathcal{B}(N_{\rm obs})$ in Eqs.~\ref{eq:t_lm_bub}, \ref{eq:cl_bub} requires averaging the one- and two-point functions of $w_{lm}$ over the distributions of collision parameters, which we now address.

Consider an observer with $N_{\rm obs}$ collisions intersecting the visible portion of the LSS.  Any one of the $N_{\rm obs}$ collision bubbles can be thought of as being selected out of a probability distribution $\mathcal{P}(z_c,\psi,\theta,\phi)$ that corresponds to the distribution over nucleation events (as discussed above, we will also assume that this suffices to determine the overall effect of all collisions).  Neglecting the effect of bubbles in bubbles (by effectively allowing bubbles in bubbles that superpose) renders each nucleation event statistically independent and so $\mathcal{P}$ should be the same for each bubble, given by normalizing $dN/dz_c d\rho d\Omega_2$ to 1.  For large $H_F/H_I$ and small $\xi_{ls}$, the 4-volume element $dV_4(z_c,\rho,\theta,\phi)$ at a given $z_c$ is very nearly independent of $\rho$, $\theta$ and $\phi$.  Thus we can write \begin{equation} \label{eq:V_42} dN(z_c, \rho, \theta, \phi)\simeq\lambda H_F^{-4} \bar{j}(z_c)d z_c d\rho d\Omega_2, \end{equation} where $\bar{j}(z_c)$ is a known function of $z_c$ which one can compute from the definition of $z_c$.  From the discussion in Sec.~\ref{sec:distributions}, $\bar{j}(z_c)$ is peaked around $z_c\sim H_F^{-1}$, and decreases for larger values of $z_c$.  Since we are concerned only with bubbles such that $\left|\rho\right|\leq1$, we change variables back to $\psi$ and normalize Eq.~\ref{eq:V_42} to yield \begin{equation} \label{eq:P} \mathcal{P}(z_c, \psi, \theta, \phi) = \frac{1}{16\pi}j(z_c) \sin\frac{\psi}{2} .\end{equation}  Here $j(z_c)$ denotes the normalized $\bar{j}(z_c)$.

The probability density $\mathcal{P}$ governs a single collision, but we can use it to average over particular realizations of multiple collision scenarios with $N_{\rm obs}$ fixed.  This is because the various quantities entering into Eq.~\ref{eq:t_lm_bub} consist of the sums of contributions from $N_{\rm obs}$ individual collisions and, if each nucleation event is statistically independent, the averages over $\left\{\mathcal{B}(N_{\rm obs})\right\}$ and the sums over bubbles commute.  Thus, the expectation values of the one- and two-point functions for many bubbles are the sums of the corresponding expectation values for individual bubbles, which are averaged over $\mathcal{P}$.  For example, the expectation value of the one-point function $w_{lm}$ for a scenario with $N_{\rm obs}$ collisions is given by \begin{equation} \label{eq:onepoint_sum} \left<w_{lm}\right>=\left<\sum_i w_{lm}^i\right>=\sum_i\left<w_{lm}^i\right>=N_{\rm obs}\left<w^0_{lm}\right>\end{equation} where $\left<w^0_{lm}\right>$ is the expectation value for a single collision, which is the same for all collisions because of statistical independence.  Similarly for the two point function, \begin{equation} \label{eq:twopoint_sum} \begin{aligned} \left<w_{l_1m_1}w^*_{l_2m_2}\right>&=\left<\sum_{i,j} w_{l_1m_1}^i w^{*j}_{l_2m_2}\right>=\sum_{i,j}\left<w_{l_1m_1}^i w^{*j}_{l_2m_2}\right>\\&=N_{\rm obs} \left<w_{l_1m_1}^0 w^{*0}_{l_2m_2}\right>\\&+N_{\rm obs}(N_{\rm obs}-1) \left<w_{l_1m_1}^0\right> \left<w^{*0}_{l_2m_2}\right> \end{aligned} .\end{equation}  We can then use the distribution $\mathcal{P}$ to perform the single-bubble averages in the last equalities of Eqs.~\ref{eq:onepoint_sum}, \ref{eq:twopoint_sum}.  The details of these calculations can be found in Appendix~\ref{sec:Ap_2}

Consider first the averages over $z_c$.  As discussed in Sec.~\ref{sec:assumptions}, the dependence of the temperature profile on $z_c$ for $H_F/H_I\gg1$, $\xi_{ls}\ll1$ arises only through $\alpha$, which additionally depends on the underlying potential and microphysics but is independent of $\psi$ in this limit.  Therefore we can absorb all of the uncertainty and model-specific information pertaining to the averages of $\alpha$ over $z_c$ into two parameters, \begin{align} \alpha_1&\equiv \left<\alpha\right>= \int \alpha(z_c) j(z_c) dz_c\\ \alpha_2&\equiv \left<\alpha^2\right>=\int \alpha^2(z_c)j(z_c)dz_c. \end{align}  Doing so allows us to treat $\alpha_{1,2}$ (along with $N_{\rm obs}$) as free parameters and work out $\left<C_l\right>$ in terms of these quantities; we defer a study of realistic ranges of $\alpha_{1,2}$ to future work.  As a consequence, our computed $\left<C_l\right>$ will describe the expected power spectrum in terms of $N_{\rm obs}$, $\alpha_{1,2}$, assuming that a scenario described by these values can occur.  Recent numerical studies \cite{Johnson:2011wt} suggest that the strength of the inflaton perturbation from a collision (and hence $\alpha$) are primarily dependent on the underlying potential, with only a mild dependence on the kinematics.  If this is the case, then for a fixed potential the averaged parameters $\alpha_1$, $\alpha_2$ should provide a good characterization of the expected signal for most observers.  Note that by Jensen's inequality, $\alpha_2\geq \alpha_1^2$.

Performing the averages over $\mathcal{P}$, the expectation value of the one-point function $\left<w_{lm}\right>$ is given by \begin{equation}  \label{eq:onepoint} \left<w_{lm}\right> =\frac{\sqrt{\pi}}{6}N_{\rm obs} \alpha_1 \xi_{ls} \delta_{l0}\delta_{m0} .\end{equation}  Only the monopole contributes to the one-point functions by the orthogonality of the spherical harmonics.  This monopole piece enters at all $l$-values by multiplying the contribution of the inflationary perturbations in Eq.~\ref{eq:t_lm_bub}.

Moving on to the two-point function, and motivated by the discussion in Appendix \ref{sec:Ap_2}, we define the quantity $w_l$ as 
\begin{align}
\label{eq:wl}
w_l=&\int\left|w_{l'0'}(\psi)\right|^2\sin\frac{\psi}{2}\delta_{ll'}d\psi  
\end{align} 
where 
\begin{equation} 
\begin{aligned} 
\label{eq:w_l0}  
w_{l'0'}(&\psi)= \int^1_{\cos\frac{\psi}{2} }Y_{l'0'}^*(\cos^{-1}x)\left(x-\cos\frac{\psi}{2}\right) dx\hspace{.5mm} \Theta\left(\pi-\psi\right) \\  
& +\int^1_{-\cos\frac{\psi}{2} }Y_{l'0'}^*(\cos^{-1}x)\left(x+\cos\frac{\psi}{2}\right) dx \hspace{.5mm} \Theta\left(\psi-\pi\right)  
\end{aligned} 
.\end{equation}  
As detailed in Appendix~\ref{sec:Ap_2}, primed indices as in Eq.~\ref{eq:w_l0} signify quantities evaluated in the polar frame, in which a collision is centered at the north pole of the celestial sphere.  The quantity $w_l$ isolates the $l$-dependence of  $\left<w_{l_1m_1}w^*_{l_2m_2}\right>$.  We plot $w_l$ for multipoles $l \leq 40$ on the left in Fig.~\ref{fig:w_l}.  In terms of $w_l$ and neglecting the monopole contribution to $a_{lm}$, the two point function $\left<a_{l_1m_1}a^*_{l_2m_2}\right>=\left<f^2\right>\left<w_{l_1m_1}w^*_{l_2m_2}\right>$ where (see Appendix B)
\begin{equation} 
\begin{aligned} 
\label{eq:twopoint} 
\left<w_{l_1m_1}\right.&\left. \hspace{-1mm} w^*_{l_2m_2}\right>=\frac{\pi^2 N_{\rm obs}\alpha_2 \xi_{ls}^2}{2l_1+1}w_{l_1} \delta_{l_1l_2}\delta_{m_1m_2} \hspace{2mm} +\\ 
&N_{\rm obs}(N_{\rm obs}-1)\left(\frac{\alpha_1 \xi_{ls} \sqrt{\pi}}{6}\right)^2 \delta_{l_10}\delta_{l_20}\delta_{m_10}\delta_{m_20} 
\end{aligned} 
.\end{equation}  
The first piece contains the expectation value of the individual two-point functions and the second piece corresponds to products of the one-point functions.  This contribution is statistically isotropic.

%For $\alpha_1 > 0$, $\left<f^2\right>\rightarrow 1$ as $\alpha \xi_{ls} \rightarrow 0$, and decreases to zero for larger $\alpha \xi_{ls}$ with $N$, corresponding to the average CMB temperature becoming much larger than its value in the absence of collisions due to the bubble hot spots.  For $\alpha < 0$, the collisions tend to cool down the CMB so that $\left<f^2\right> > 1$ and has a singularity at $N\alpha \xi_{ls}/6\sim1$.  This singularity corresponds to $\left<T_0^2\right>\rightarrow 0$, in which case the cold spots from the collisions drive the average CMB temperature squared to $0$.  One should already be suspect of scenarios for which $N\alpha \xi_{ls}/6 \gtrsim 1$, since in this case the temperature``perturbations" are larger than the background temperature itself; to avoid these issues we restrict our attention to physically tenable scenarios with $N\alpha \xi_{ls}/6 \lesssim 1$.  

It remains to average the Gaunt coefficient piece of Eq.~\ref{eq:t_lm_bub} as well as to compute the quantity $\left<f^2\right>$ .  Considering the last term in Eq.~\ref{eq:t_lm_bub}, we define \begin{equation} \begin{aligned} \label{eq:Rl} R_l=& \sum_m\sum_{l_1,m_1,l_2,m_2}\frac{w_{l_1}}{2l_1+1} \left(R_{lm}^{l_1m_1l_2m_2}\right)^2 C_{l_2}^{bb} \end{aligned} .\end{equation}  We plot $R_l$ on the right in Fig.~\ref{fig:w_l} along with $C_l^{bb}$ for reference.  Clearly, $R_l \ll w_l$ since by Eq.~\ref{eq:Rl} it involves the product of perturbation terms $w_l C_l^{bb}$, and the Gaunt coefficients are typically $\mathcal{O}(10^{-1})$ or smaller in our computations.  In addition to $R_l$, there is also a contribution $\pi^2\alpha_1^2 \xi_{ls}^2 R_l^0(N_{\rm obs})$ to $C_l$ where \begin{equation} \label {eq:Rl0} R^0_l(N_{\rm obs}) = \frac{N_{\rm obs}(N_{\rm obs}-1)}{144 \pi^2}C_l^{bb} .\end{equation}  This term arises from the second line of Eq.~\ref{eq:twopoint} when inserted into Eq.~\ref{eq:t_lm_bub}.    

To compute $\left<f^2\right>$ we must take the ensemble average of the inverse of the sky-averaged temperature.  We provide an explicit integral expression for $\left<f^2\right>$ in Appendix~\ref{sec:Ap_2} which depends on $N_{\rm obs}$, $j(z_c)$, and the unknown function $\alpha(z_c)$.  This contribution will simply result in an overall scaling of $\left<C_l\right>$.  A lower bound on this quantity can be derived by Jensen's inequality (see Appendix~\ref{sec:Ap_2}), which yields $\left<f^2\right>\gtrsim 1/(1+N_{\rm obs}\alpha_1 \xi_{ls}/6)$.  The behavior of $\left<f^2\right>$ will generally depend on the sign of $\alpha_1$, since a multiple collision scenario producing mostly cold spots on the CMB will tend to lower $T_0$ with respect to $T_0'$, driving $\left<f^2\right>>1$ while the opposite is true if $\alpha_1>0$.  Regardless, we expect $\left<f^2\right>\approx1$ in the perturbative regime with which we are concerned, since $N_{\rm obs} \alpha_1 \xi_{ls}\ll1$ in this case.

We are now in position to evaluate $\left<C_l\right>$ in terms of the relevant parameters of the theory.  Using Eqs.~\ref{eq:multi_prof}, \ref{eq:cl_def}, \ref{eq:onepoint}, \ref{eq:twopoint}, \ref{eq:Rl}, \ref{eq:Rl0} and performing the sums involving the Gaunt coefficients, we arrive at: 
\begin{widetext} \begin{equation} \begin{aligned} \label{eq:spectrum} \left<C_l\right>=&\left<f^2\right>\left[\frac{\pi^2N_{\rm obs} \alpha_2 \xi_{ls}^2}{2l+1} \left(w_l+R_l\right) +C_l^{bb}\left(1+\frac{N_{\rm obs} \alpha_1 \xi_{ls}}{6}+\frac{N_{\rm obs}(N_{\rm obs}-1)\alpha_1^2\xi_{ls}^2}{144}\right)\right] \end{aligned} .\end{equation} 
\end{widetext} (Note that we have dropped the $l=0$ piece from Eq.~(\ref{eq:w_lm}) and written out $R_l^0(N_{\rm obs})$ explicitly as the last term.)  This is our main result, and characterizes the effects of multiple collisions on the CMB power spectrum under our assumptions.  The expected angular power spectrum depends only on three parameters, $N_{\rm obs}$, $\alpha_1$, and $\alpha_2$ which encode the model-specific information.  The shape of the power spectrum itself results from the isotropic and scale-invariant distribution of bubbles which should apply regardless of the model potential, as long as $H_F/H_I\gg1$ and $\xi_{ls}\ll1$.  

As one might expect, the most significant contribution to $\left<C_l\right>$ is that of $w_l$.  From Fig.~\ref{fig:w_l}, $R_l\ll w_l$ so its contribution is negligible.  There is a scale-invariant contribution in Eq.~\ref{eq:spectrum} proportional to $N_{\rm obs} \alpha_1$, and so one might imagine it becoming comparable to that of $w_l$ for small $\alpha_{1,2}$.  However, this occurs only for very small values of $\alpha_2$ such that \begin{equation} \label{eq:scale_invariant} \alpha_2 \xi_{ls}^2 \lesssim \frac{(2l+1)^2}{36\pi^4} \left(\frac{C_l^{bb}}{w_l}\right)^2 .\end{equation}  For example, for the scale invariant piece to be comparable to that of $w_l$ for $l\geq 10$ requires $\alpha_2 \lesssim 10^{-11}$ by Eq.~\ref{eq:scale_invariant}, assuming $\xi_{ls}=.05$.  For it to dominate at lower $l$ would require even smaller $\alpha_2$, in which case the effects of the collisions would be very difficult to discern except for very large $N_{\rm obs}$.  Even if it were there, this contribution would be degenerate with a slightly different power in the primordial perturbations, at least for the 2-point function.

%%%%%%%%%%%%%%%%%%%%%%%%%%%%%%%%%%%%%%%%%%%%%%%%%%%
\begin{figure*}[!t]
\mbox{\hspace{-1cm}\includegraphics[width=0.6\textwidth,clip]{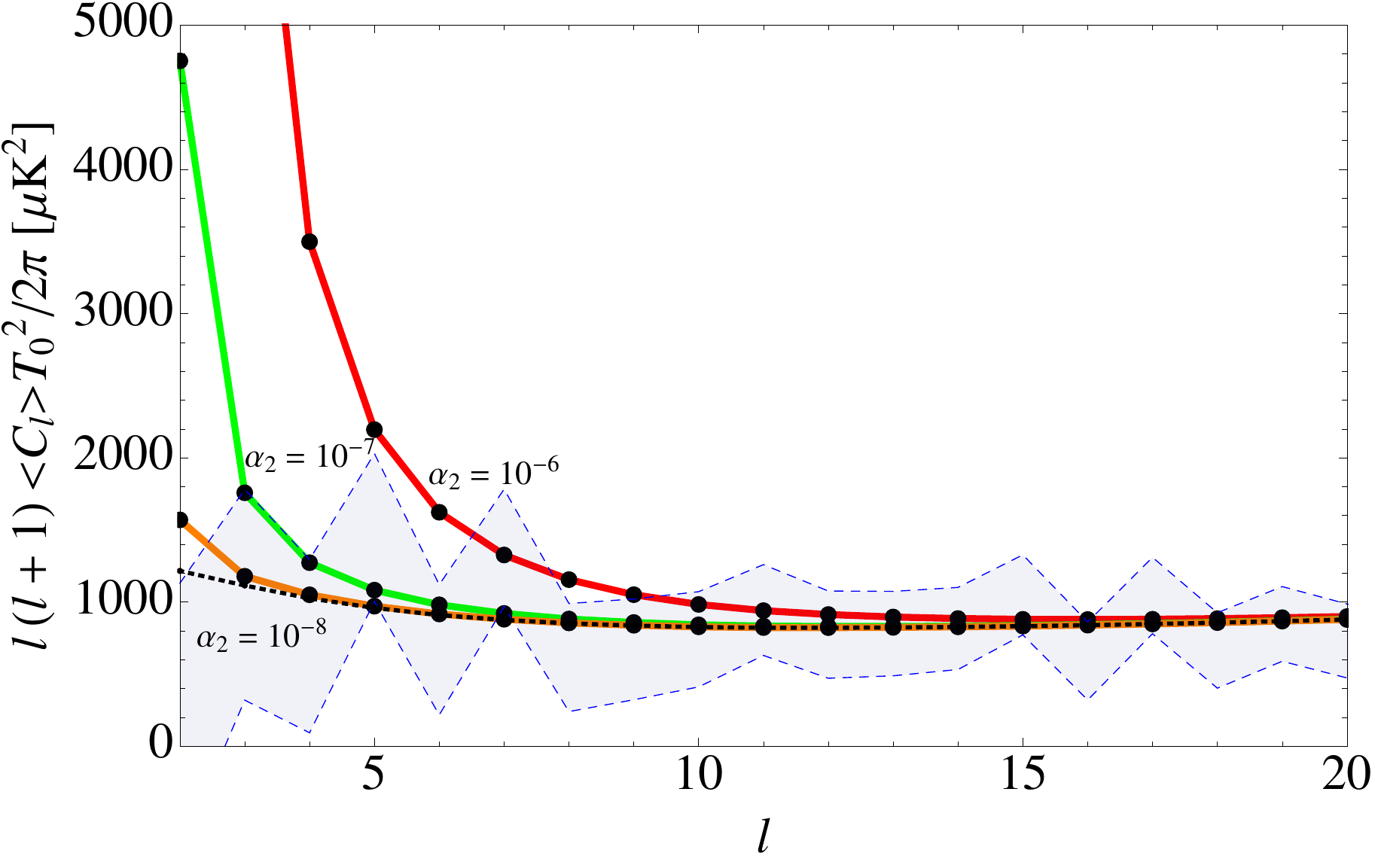}}
\caption{\label{fig:Cl}\small Expected temperature anisotropy angular power spectrum for multiple collision scenarios from Eq.~\ref{eq:spectrum_approx} with $N_{\rm obs}=100$, $\xi_{ls}=.05$, and for $\alpha_2=10^{-6},10^{-7},10^{-8}$ in red, green, and orange, respectively.  The $C_l^{bb}$ are shown by the dotted black line for reference.  Points in the shaded blue region lie within the approximate WMAP7 error bars \cite{WMAP}.  The increase in power on large angular scales is apparent.}
\end{figure*}
%%%%%%%%%%%%%%%%%%%%%%%%%%%%%%%%%%%%%%%%%%%%%%%%%%%

Using this fact, along with $\left<f^2\right>\approx 1$, the expected power spectrum describing a multiple collision scenario simplifies to \begin{equation} \label{eq:spectrum_approx} \left<C_l\right>\simeq \frac{\pi^2N_{\rm obs} \alpha_2 \xi_{ls}^2}{2l+1} w_l +C_l^{bb} .\end{equation}  This spectrum is plotted in Fig.~\ref{fig:Cl} for $N_{\rm obs}=100$ and various values of $\alpha_2$, along with the approximate error bars from the WMAP 7 year results \cite{WMAP}.  In this scenario, the effective power spectrum depends only on $\alpha_2$ and $N_{\rm obs}$.  From Eq.~\ref{eq:twopoint}, the collisions will not contribute any statistical anisotropy to the expected spectrum.

An important feature of the resulting spectrum in Fig.~\ref{fig:Cl} is the dramatic increase in power of the low multipoles.  This behavior can already be seen in the quantity $w_l$ as in Fig.~\ref{fig:w_l} and has a physical origin: the flat probability distribution over bubble scales yields an equal number of large and small bubbles, and the bubble profile is ``smooth" so that large bubbles contribute only to relatively low-$l$ multipoles. Yet larger bubbles both take up a larger fraction of the sky (per bubble), and have a larger central amplitude. (The larger the affected disk on the sky, the farther out on the affected portion of the last scattering surface -- and hence the closer to the collision event -- the observer can see; this is evident in the temperature profile Eq.~(\ref{eq:multi_prof}), since the maximum amplitude of the temperature perturbation for a given collision scales as $\left|T_{max}\right|\sim \left|\alpha_i\right| \xi_{ls} (1-\cos \psi_i/2)$.)  Since larger bubbles tend to be brighter, searches for collision disks on the CMB temperature map, such as those of Ref.~\cite{Feeney:2010dd}, may still be able to discern several distinct disks, corresponding to the brightest collisions, as long as $N_{\rm obs}$ is not too large.   

The peak in power occurs regardless of the sign of $\alpha(z_c)$, since this effect arises from terms proportional to $\alpha_2$.  For there to be a decrease in power, the scale-invariant contribution to Eq.~\ref{eq:spectrum} would have to be large with $\alpha_1<0$.  Even if this were the case, from Eq.~\ref{eq:f_approx} the product with $\left<f^2\right>$ will still tend to increase $\left<C_l\right>$ with respect to $C_l^{bb}$. 

As $N_{\rm obs}$ becomes large, the observed spectrum will tend to converge to its expectation value Eq.~\ref{eq:spectrum}, providing an increasingly sharp prediction for a given multiple collision scenario.  This is simply a statement of Bernoulli's Theorem.  Consider, for example, the one point function: neglecting bubbles within bubbles renders the individual $w^i_{lm}$ statistically independent and a multiple collision scenario can then be thought of as nucleating $N_{\rm obs}$ bubbles according to the distribution $\mathcal{P}$.  In this case, Bernoulli's Theorem suggests that with more successive nucleation events, the mean value of the various $w^i_{lm}$ should converge to the expectation value $\left<w^0_{lm}\right>$, i.e. \begin{equation} \label{eq:LOLN} \frac{1}{N_{\rm obs}}\sum w^i_{lm}\rightarrow \left<w^0_{lm}\right>\end{equation} as $N_{\rm obs}\rightarrow \infty$.  Multiplying through by $N_{\rm obs}$, the sum on the LHS of Eq.~\ref{eq:LOLN} is simply a particular realization of a multiple collision scenario, while the quantity on the RHS is the expectation value $\left<w_{lm}\right>$ (c.f. Eq.~\ref{eq:onepoint_sum}).  Thus, the expectation value $\left<w_{lm}\right>$ should describe the observed $w_{lm}$ with increasing accuracy for large $N_{\rm obs}$.  Similar arguments apply to the two-point functions.  Therefore, for large $N_{\rm obs}$, the observer should expect to see a spectrum very close to $\left<C_l\right>$.  To quantify this assertion and investigate the theoretical uncertainty of Eq.~\ref{eq:spectrum} across realizations of multiple collision scenarios, one might imagine computing the variance of the $\left<C_l\right>$.  However, doing so would require knowledge of the distribution of $\alpha(z_c)$ which we have not attempted to compute and so we do not perform this analysis here.  

Whether or not scenarios with $N_{\rm obs}\gg1$ are plausible depends on the underlying potential and microphysics.  From the discussion surrounding Eq.~\ref{eq:Nobs}, there would seem to be cases where a large $N_{\rm obs}$ is quite natural.  In treating $N_{\rm obs}$ as a free parameter, we have avoided the subtleties of this issue and expect the profile Eq.~\ref{eq:spectrum} to provide a good description of the effect of multiple collisions on the CMB spectrum for large $N_{\rm obs}$.  

\subsection{CMB Constraints On Multiple Collisions: A First Look}

Using the results of Eqs.~\ref{eq:spectrum}, \ref{eq:spectrum_approx}, we can preliminarily assess how scenarios with $N_{\rm obs}$ fare in light of observations of the CMB temperature spectrum.  Of course constraining models based on only one set of observations is difficult and a much more in-depth study is required to determine how likely it is that we live in a multiple collision scenario.  We simply wish to answer the question: if multiple collisions have impacted our bubble producing precisely the expected spectrum, Eq.~\ref{eq:spectrum_approx}, what do observations of the CMB data tell us about the parameters $N_{\rm obs}$ and $\alpha_2$?  

From Fig.~\ref{fig:Cl}, we expect that the strongest constraints on $N_{\rm obs}>1$ in our approximation\footnote{We stress that our analysis applies only to angular scales $\gtrsim 1^{\circ}$, corresponding roughly to $l\lesssim 70$.  To determine the effects of collisions on smaller scales requires going beyond the Sachs-Wolfe approximation, which we defer to future work.} will arise from observations of the CMB quadrupole, since its observed value is low compared to the best-fit $\Lambda$CDM cosmology and multiple collisions will tend to exacerbate this discrepancy.  A detailed statistical analysis in Ref.~\cite{WMAP_quad} has concluded that the best fit $\Lambda$CDM quadrupole is in fact well within the $95\%$ C.L. of the WMAP7 observed value.  Comparing $\left<C_2\right>$ with the $95\%$ C.L. upper bound on $C_2^{obs}$, we find that the parameters $N_{\rm obs}$ and $\alpha_2$ should satisfy \begin{equation} \label{eq:quadrupole} N_{\rm obs} \alpha_2 \Omega_c \lesssim3.5\times 10^{-9} .\end{equation}  Note that this also constrains the magnitude of $\alpha_1$, since $\alpha_1^2\leq \alpha_2$, but not its sign.  

The observed CMB dipole does not constrain the parameters $N_{\rm obs}$, $\alpha_2$ as strongly as the quadrupole.  If we imagine that all of the observed dipole has arisen from the intrinsic perturbation to the gravitational potential from collisions, consistency with the observed value of the dipole dictates $N_{\rm obs}\alpha_2 \Omega_c \lesssim 1.5\times 10^{-6}$.  Without taking the frame shift into account for $\left|\rho\right|\leq1$ bubbles, the bound would be slightly strengthened, $N_{\rm obs}\alpha_2\Omega_c \lesssim 2.5\times 10^{-7}$, but still not competitive with that provided by the quadrupole.

%Of course how seriously one should take these estimates depends on the size of $N_{\rm obs}$ and the distribution of $\alpha(z_c)$.  These bounds assume that the observer sees a bubble collision spectrum exactly representative of the expectation values.  By the discussion in Sec.~\ref{subsec:averaging}, we expect this to be a good description for large $N_{\rm obs}$.  Precisely how large $N_{\rm obs}$ must be for this statistical picture to be satisfactory depends in particular on $\alpha(z_c)$ and we leave this consideration to future study.  Our derived bounds on the collision parameters should be taken simply as an illustrative estimate of the range of how mild the average collision must be for a given $N_{\rm obs}$ to be consistent with CMB observations.  Also, whether or not scenarios satisfying Eq.~\ref{eq:quadrupole} are feasible remains to be investigated.

A few notes on this result are in order:
\begin{itemize}
\item This result is, of course, dependent upon the detailed assumptions explained in previous sections - namely, the radial profile of a single bubble, the flat distribution in bubble angular sizes, the ``frame shift", and the approximation that bubbles superpose.  However, all of these are unlikely to be dramatically incorrect in a realistic scenario.
\item It is also not necessarily the case that the CMB temperature power spectrum is the best observational constraint on multiple collision scenarios.  Observations of e.g. large scale structure, as well as CMB polarization and non-Gaussianity, will also be affected by the collisions and should be investigated.  These signatures are left for future extensions of this work.
\item Our results are useful primarily for large $N_{\rm obs}$; for a small number of bubbles, searching for disk templates on the sky should provide much stronger constraints.
\end{itemize}

\section{Summary And Conclusions}\label{sec:conclusion}

In this study, we have begun to address the possibility of having many cosmic bubble collisions influencing the visible portion of our last scattering surface.  This can occur despite vanishingly small nucleation rates as long as it is compensated for by a large ratio of Hubble constants between the false vacuum and the presumed inflationary phase in the observation bubble.  For such a scenario to be compatible with the existence of observers (and to not produce effects clearly inconsistent with observation), the effects of the collisions must be mild, suggesting that one should be able to treat the effects of collisions perturbatively for models of interest.  We were thus led to consider a scenario where many bubbles collide with the observation bubble while leaving the interior FRW foliation intact, supporting inflation and a standard background cosmology to the future of the collisions and allowing us to study the effects of the collisions on the last scattering surface and CMB temperature spectrum. 

Given a number of reasonable simplifying assumptions, we are led to a model for these perturbations with the following characteristics.
\begin{itemize}

\item The overall effect on the last-scattering surface and on the CMB temperature is necessarily small (assumed for compatibility with observations.)

\item The total perturbation on the last-scattering surface is a superposition of individual bubble perturbations (see Sec.~\ref{subsec:intervening}) resulting from collisions with bubbles decaying to the same vacuum (which may or may not match that of the observation bubble.)
	
	\item Each bubble appears to cover less than half of the sky (see Sec.~\ref{subsec:frame}), with a size distribution that is flat in $\cos\psi$, where $\psi$ is the angular size of the perturbed disk (see Sec.~\ref{subsec:prob}).
	
	\item Each bubble's profile is a `truncated dipole', with central amplitude dependent on $\psi$ and $z_c$, a `kinematic' parameter describing the (dS-invariant) distance between the bubbles' nucleation points (see Sec.~\ref{subsec:single}).
	 
	\item The details of the kinematics and the inflaton potential that determine the perturbation strength can be folded into two parameters $\alpha_{1,2}$.  The other relevant quantity is the total number $N_{\rm obs}$ of (less than full-sky) bubbles on the last-scattering surface.  These three parameters are sufficient to describe the effects of many collisions on the CMB temperature.

\end{itemize}

We put this picture forward as a scenario that is both sufficiently simple that predictions can be computed in detail from it, and including enough aspects of a realistic scenario so that it should give an accurate overall picture of the expected effect of a many-bubble scenario.

The above picture, to a good approximation, yields a contribution to the $l \lesssim 70$ power spectrum of 
\begin{equation}
\left<C_l\right>\simeq \frac{\pi^2N_{\rm obs} \alpha_2 \xi_{ls}^2}{2l+1} w_l,
\end{equation} 
where $\xi_{ls} \sim 2\sqrt{\Omega_c}$, $\alpha_2$ encodes features of the inflaton potential that determine the amplitude of a given collision perturbation, and $w_l$ is a fixed function given by Eq.~\ref{eq:wl} and shown in Fig.~\ref{fig:w_l}.  This equation holds with increasing accuracy as $N_{\rm obs}$ becomes large.  The resulting power spectrum contribution is very red-tilted (see Fig.~\ref{fig:Cl}), so that the strongest limit on a multi-bubble scenario comes from the quadrupole.   

  In our universe, a multiple collision scenario with $N_{\rm obs}$ visible bubbles and average brightness described by $\alpha_2$ consistent with observations of the CMB power spectrum would require roughly $\alpha_2 \Omega_c \lesssim 3.5\times 10^{-9}/N_{\rm obs}$. 
    This bound can in turn be used to constrain a given well-specified scenario by developing in detail the link between the inflaton potential and the parameters $\alpha_2,$ $\Omega_c$, and $N_{\rm obs}$.  
  
  We emphasize that this is a first study aimed at laying the groundwork for such scenarios and computing constraints using the power spectrum.  It is quite possible, and even likely, that other observables like polarization or non-Gaussianity could give significantly stronger constraints, or provide a cleaner method for the actual detection of bubble effects.  This set of methods comprises a complement for those already developed to investigate scenarios with a few bubble impacts.  In either case, refinement of these tests will soon provide real constraints on the parameter space of eternal inflation, and might even produce evidence for eternal inflation itself.

\begin{acknowledgments}
\noindent  We thank Matt Johnson for helpful conversations.  AA is supported by NSF Grant PHY-0757911 and by a ``Foundational Questions in Physics and Cosmology" grant from the John Templeton Foundation. JK is partly supported by an Outstanding Junior Investigator Award from the US Department of Energy and by Contract DE-FG02-04ER41268, and by NSF Grant PHY-0757911.
\end{acknowledgments}

\appendix

\addappheadtotoc

\section{Intervening Bubbles} \label{sec:Ap_1}

In this Appendix we detail the calculation of the expected number of intervening bubbles to intersect a colliding bubble with nucleation center $(z_c, \rho)$.  Since we are interested in intervening bubbles intercepting late-time collision bubbles, we can neglect the angular dependence of the initial value surface and choose $\theta_c=\phi_c=0$ without loss of generality.  

By the discussion in Sec.~\ref{subsec:intervening}, we must calculate the 4-volume $\mathcal{I}(z_c,\rho)$ with the cosmologies fixed.  $\mathcal{I}(z_c,\rho)$ is given by \begin{equation} \begin{aligned} \label{eq:I} \mathcal{I}(z_c,\rho)=&V_4^{PLC}[\eta_0(z_c,\rho),T_0(z_c,\rho)]\\&-V_4^{PLC}[\eta_c(z_c,\rho),T_c(z_c,\rho)]-V_4^{PLC}[0,0] \end{aligned} \end{equation} where $V_4^{PLC}[\eta,T]$ denotes the 4-volume in the past light cone ($PLC$) from the point $(\eta,T)$ and $(\eta_0,T_0)$ is the point of intersection of the future light cone from $(\eta_c(z_c,\rho),T_c(z_c,\rho))$ and the observation bubble wall.  

It remains to determine whether a point $x_2=(\eta_2,T_2,\theta_2,\phi_2)$ is within the past light cone of $x_1=(\eta_1,T_1,\theta_1,\phi_1)$.  In the embedding space, two points $X^{\mu}_1$, $X^\mu_{2}$ are timelike or null separated if and only if \begin{equation} \eta_{\mu\nu} X_1^{\mu}X_2^{\nu} \geq 1 .\end{equation}  Restricting this relation to the dS hyperboloid gives an equivalent relation \begin{equation} \begin{aligned}P(x_1,x_2)\equiv&-\tan T_1 \tan T_2 + \frac{\sin\eta_1 \sin\eta_2}{\cos T_1 \cos T_2}\omega_i(x_1)\omega_j(x_2)\delta_{ij}\\ & +\frac{\cos\eta_1\cos\eta_2}{\cos T_1 \cos T_2}\geq 1 \end{aligned} \end{equation} for timelike or null separated points, where the $\omega_i$ are defined as in Eq.~\ref{eq:embedding}.  We can then numerically integrate over the 4-volume to the past of each of the points $x_i$ in Eq.~\ref{eq:I}, weighted by the step function $\Theta(P(x_i,x)-1)$ to obtain $\mathcal{I}(z_c,\rho)$.  The integration within the set of points $PLC[0,0]$ can be done analytically, however the angular dependence of the lightcones away from the origin makes it difficult to do so for the other points.  We cut off the numerical integration at $T$ corresponding to the intersection of the PLC from $(\eta_c,T_c)$ and that from the origin.  This does not count regions near past infinity (corresponding to early time bubbles), however these portions of the false vacuum are typically included in $PLC[\eta_c,T_c]\cap PLC[0,0]$ and so will not contribute to $N_{int}$. 

To obtain $N_{\rm int}$ we use Eq.~\ref{eq:N_int} with a particular choice of $\lambda$.  In Fig.~\ref{fig:lambda_min} we use $\lambda$ such that $N_{\rm obs}=1$ from Eq.~\ref{eq:Nobs}.  Increasing $N_{\rm obs}$ will correspond to increasing either $\lambda$ or $\xi_{ls}$, however we still find $N_{\rm int}\ll1$ for cases of interest.

\section{Calculating Expectation Values Of Correlation Functions}  \label{sec:Ap_2}

In this Appendix we detail the calculations of the expected one- and two-point correlation functions Eqs.~\ref{eq:onepoint}, \ref{eq:twopoint}.  Our conventions follow that of Ref.~\cite{book}, to which we refer the reader for further details as well as useful identities used in the calculations we present.

We are interested in the two-point function of the temperature anisotropy which in turn will involve the one- and two-point functions of $w_{lm}$.  Our analysis is simplified by the following observation: for a given function $g$ expanded in spherical harmonics in a coordinate system (primed indices), the coefficients in a rotated coordinate system (unprimed indices) are given by \cite{book, Gordon:2005ai} \begin{equation} \label{eq:rotate} g_{lm}=\sum_{m'} g_{l'm'}D^l_{m'm}(\gamma,\theta,\phi)\delta_{ll'} \end{equation} where $\gamma,\theta,\phi$ are the Euler angles specifying the transformation between the two coordinate systems and $D^l_{m'm}$ is the Wigner rotation matrix.  Since $w_{lm}$ is a sum of the coefficients corresponding to the individual bubbles, we can write $w_{lm}=\sum_{N_1+N_2} w_{lm}^i$ and evaluate each $w_{lm}^i$ in a frame in which the collision is centered at the north pole $\theta=\phi=0$ and rotate them back into the original coordinate system via the transformation Eq.~\ref{eq:rotate}.  

Since, in the polar (primed) frame, the profile for each bubble on the sky by construction is azimuthally symmetric about the z-axis, only the $m'=0$ modes contribute and we are left with  \begin{equation}\label{eq:w_lm} w_{lm}=\sum_{i} w_{l'0'}^iD_{0'm}^l\left(\alpha_i,\beta_i,\gamma_i\right) \delta_{ll'} \end{equation} where \begin{gather} w_{l'm'}^i=\int Y_{l'm'}^*(\theta,\phi)w^i(\hat{\mathbf{n}})|_{\theta_{0i}=0,\phi_{0i}=0} \hspace{1mm} d\Omega\\  \label{eq:rotation} D_{0'm}^l\left(\alpha_i,\beta_i,\gamma_i\right)=(-1)^m\sqrt{\frac{4\pi}{2l+1}}Y_{lm}^*(\beta_i,\gamma_i) \end{gather}  In our calculations it will often be convenient to consider the quantity $w_{l'0'}(\psi)$, corresponding to the coefficient $w_{lm}$ from a collision centered at the north pole $\theta=\phi=0$ (hence the prime), with angular scale $\psi$, and with the strength $\alpha \xi_{ls}$ factored out: \begin{equation} \label{eq:wl0} w_{l'0'}(\psi_i)\equiv \frac{1}{2\pi \alpha_i \xi_{ls}}w_{l'0'}^i .\end{equation}  This is precisely the quantity explicitly defined in Eq.~\ref{eq:w_l0}.  The relevant Euler angles to rotate from the polar frame to the original frame are given in our convention by $\beta_i=\theta_{0i}$, $\gamma_i=\pi-\phi_{0i}$ ($\beta_i=\pi-\theta_{0i}$, $\gamma_i=\pi+\phi_{0i}$) for $i\leq N_1$ ($i>N_1$).  The coefficients $a_{lm}$ are related to $w_{lm}$ via $a_{lm}=f w_{lm}$ where we have neglected the monopole piece arising from the $l=0$ integral.

To carry out the ensemble averages for the various correlation functions, we average by the density $\mathcal{P}$ defined in Eq.~\ref{eq:P}.  We use the shorthand \begin{equation} \label{eq:distap} g(\psi)=\frac{1}{16\pi}\sin\frac{\psi}{2} \end{equation} for brevity.  In this prescription, the expected one point function is given by \begin{widetext} \begin{equation} \begin{aligned} \label{eq:ap1}  \left<w_{lm}\right>&= \left<\sum_iw_{lm}^i\right>\\ &= (-1)^m\sqrt{\frac{16\pi^3 \xi_{ls}^2}{2l+1}}\left<\alpha(z_c)\right>\left(N_1 \left<Y_{lm}^*(\theta,\pi-\phi)\right>\left<w_{l'0'}(\psi)\right>_{\psi \leq \pi}+N_2  \left<Y_{lm}^*(\pi-\theta,\pi+\phi)\right>\left<w_{l'0'}(-\psi)\right>_{\psi \geq \pi}\right)\delta_{ll'}\\&=N_{\rm obs}\alpha_1 \xi_{ls}(-1)^m\sqrt{\frac{16\pi^3}{2l+1}}\delta_{ll'}\int w_{l'0'}(\psi)g(\psi)d\psi  \int Y_{lm}^*(\theta,\pi-\phi)d\Omega \\ &=\frac{8\pi^2 N_{\rm obs}\alpha_1 \xi_{ls} (-1)^m}{\sqrt{2l+1}}\delta_{ll'}\delta_{l0}\delta_{m0} \int w_{l'0'}(\psi)g(\psi)d\psi \end{aligned} .\end{equation} \end{widetext}  Here the brackets $\left< \right>_{\psi\leq \pi (\geq \pi)}$ denote averaging over $\psi$ in the interval $\left[0,\pi\right]$ ($\left[\pi,2\pi\right]$).  In going from the first to second line, we use Eqs. \ref{eq:w_lm}, \ref{eq:rotation}, and \ref{eq:wl0} to express the sum as a collection of averages in the polar frame with the appropriate rotation coefficients.  In going from the second to the third line we use the statistical independence of the collisions write out the averages explicitly in terms of the averages for an individual collision (as per Eq.~\ref{eq:onepoint_sum}) and use the fact that $\int Y^*_{l,m}(\theta,\pi-\phi)d\Omega=\int Y^*_{lm}(\pi-\theta,\pi+\phi) d \Omega$ to combine the $N_1$ and $N_2$ terms.  The last line follows from integrating the spherical harmonic.  

Only the monopole part contributes to the one-point function and we can perform the integrals over $\psi$, using Eq.~(\ref{eq:w_l0}) to express $w_{0'0'}$ as \begin{equation} \begin{aligned} w_{0'0'}(\psi)=&\frac{1}{\sqrt{\pi}}\left[\sin^4\frac{\psi}{4}\Theta\left(\pi-\psi\right)+\cos^4\frac{\psi}{4}\Theta\left(\psi-\pi\right) \right] \end{aligned} .\end{equation}  Inserting this into the last line of \ref{eq:ap1} and integrating yields Eq.~(\ref{eq:onepoint}).

%which yields a factor of $2/3\sqrt{\pi}\alpha_1 \xi_{ls}$.  Inserting this into the last line of \ref{eq:ap1} yields Eq.~(\ref{eq:onepoint}).

Moving on to the expectation value of the two point function, $\left<w_{l_1m_1}w_{l_2m_2}^*\right>$, we compute: \begin{widetext}\begin{equation} \begin{aligned} \left<w_{l_1m_1}w_{l_2m_2}^*\right> &= \left(-1\right)^{m_1+m_2}\frac{16\pi^3 \xi_{ls}^2 \delta_{l_1l_1'} \delta_{l_2l_2'}}{\sqrt{(2l_1+1)(2l_2+1)}} \left<\sum_{i,j}\alpha_i \alpha_j w_{l'_10'}(\psi_i)Y_{l_1m_1}^*(\theta_i,\phi_i)w_{l'_20'}(\psi_j)Y_{l_2m_2}(\theta_j,\phi_j)\right>  \\ &= \left(-1\right)^{m_1+m_2}\frac{16\pi^3 \xi_{ls}^2 \delta_{l_1l_1'} \delta_{l_2l_2'}}{\sqrt{(2l_1+1)(2l_2+1)}}\left( \sum_{i=j}\left<\alpha(z_c)^2\right>\left<w_{l_1'0'}(\psi)w_{l_2'0'}(\psi)\right> \left<Y_{l_1m_1}^*Y_{l_2m_2}\right> \right. \\ & \left. +\sum_{i \neq j} \left<\alpha(z_c)\right>^2\left<w_{l_1'0'}(\psi_i)w_{l_2'0'}(\psi_j)\right>\left<Y_{l_1m_1}^*(\theta_i,\pi-\phi_i)Y_{l_2m_2}(\theta_j,\pi-\phi_j)\right>\right) \\ & = \frac{16\pi^3\xi_{ls}^2\delta_{l_1l_1'} \delta_{l_2l_2'}}{\sqrt{(2l_1+1)(2l_2+1)}}\Big( N_{\rm obs}\alpha_2 \int\left|w_{l_1'0'}(\psi)\right|^2 g(\psi)d\psi \delta_{l_1l_2}\delta_{m_1m_2} \\ & +N_{\rm obs}(N_{\rm obs}-1)\alpha_1^2 \left<w_{l_1'0'}\right>\left<w_{l_2'0'}\right>\left<Y^*_{l_1m_1}\right>\left<Y_{l_2m_2}\right>\Big) \\ &=\frac{\pi^2 N_{\rm obs} \alpha_2 \xi_{ls}^2}{2l_1+1}w_{l_1}\delta_{l_1l_2}\delta_{m_1m_2}+N_{\rm obs}(N_{\rm obs}-1)\left(\frac{\alpha_1 \xi_{ls} \sqrt{\pi}}{6}\right)^2 \delta_{l_10}\delta_{l_20}\delta_{m_10}\delta_{m_20} \end{aligned}  .\end{equation} \end{widetext}  In the first line, we write out the average as a collection of $N_{\rm obs}$ collisions in the polar frame with the appropriate rotation coefficients.  Note that the function $w_{l'0'}(\psi)$ is real-valued, hence only the complex conjugate of the spherical harmonic appears here.  The second line follows from the commutation of the averages.  In the third equality, we use the statistical independence of the individual collisions to write out the average two-point function explicitly in terms of the averages for an individual collision and integrate over $\alpha(z_c)$.  We don't write out the one-point averages explicitly for brevity; they are precisely as in Eq.~\ref{eq:ap1}.  Between the third and fourth equality we perform the remaining integrals, again combining the $N_1$ and $N_2$ terms, since $\left< Y_{l_1m_1}^*(\theta,\pi-\phi)Y_{l_2m_2}(\theta,\pi-\phi)\right>=\left< Y_{l_1m_1}^*(\pi-\theta,\pi+\phi)Y_{l_2m_2}(\pi-\theta,\pi+\phi)\right>$ and similarly for the one-point averages.

Finally, we can compute $\left<f^2\right>$.  The quantity $f$ relates the average CMB temperature to the average temperature without any collisions.  One must then average the spatial temperature profile over the sky, then take the ensemble average.  Let us first consider the quantity $f^{-1}$ and write \begin{equation} \begin{split} \frac{T_0}{T'_0}=\frac{1}{T'_0}\left<T(\hat{\mathbf{n}})\right>_{sky}  = 1+\sum_i \left< w^i(\hat{\mathbf{n}})\right>_{sky} \end{split} .\end{equation}  The sky average for a given $w_i$ is independent of the collision's location on the sky and thus depends only on $\psi_i$.  We can thus define the $\alpha$-independent quantity \begin{equation} \mathcal{S}(\psi)\equiv\frac{1}{4\pi \alpha \xi_{ls}} \int w^i(\hat{\mathbf{n}})|_{\theta_{0i}=0,\phi_{0i}=0} \hspace{1mm} d\Omega =\frac{1}{2\sqrt{\pi}} w_{0'0'}(\psi) \end{equation} in terms of which, $f^2$ is given by  \begin{equation} f^2=\left(1+\xi_{ls} \sum_i\alpha_i\mathcal{S}(\psi_i)\right)^{-2} .\end{equation}  Using these definitions, squaring, and averaging by the distribution Eq.~\ref{eq:distap}, we arrive at our expression for $\left<f^2\right>$ in terms of $N_{\rm obs}$, $\alpha(z_c)$, and $\xi_{ls}$: \begin{widetext} \begin{equation} \left<f^2\right>=\int\frac{(4\pi)^2 g(\psi_1)g(\psi_2)j(z_{c1})j(z_{c2}) d\psi_1d\psi_2 dz_{c1}dz_{c2} }{1+N_{\rm obs} \xi_{ls}\left(\alpha(z_{c1})\mathcal{S}(\psi_1)+\alpha(z_{c2})\mathcal{S}(\psi_2)\right)+N_{\rm obs}(\alpha(z_{c1}) \xi_{ls})^2\mathcal{S}^2(\psi_1)+N_{\rm obs}(N_{\rm obs}-1)\alpha(z_{c1})\alpha(z_{c2}) \xi_{ls}^2\mathcal{S}(\psi_1)\mathcal{S}(\psi_2)} .\end{equation} \end{widetext} 

A useful lower bound on $\left<f^2\right>$ can be derived from Jensen's inequality, whereby $\left<f^2\right>\geq\left<1/f\right>^{-2}$.  Dropping the term quadratic in $N\alpha_1 \xi_{ls}$ results in  \begin{equation} \label{eq:f_approx} \left<f^2\right>\gtrsim \left(1+\frac{N_{\rm obs}\alpha_1 \xi_{ls}}{6}\right)^{-1} .\end{equation}  In our perturbative approach we expect $\left<f^2\right>\approx 1$.  This quantity simply results in an overall scaling of the power spectrum.


\begin{thebibliography}{300}

\bibitem{Aguirre:2007gy} 
  A.~Aguirre,
  %``Eternal Inflation, past and future,''
  arXiv:0712.0571 [hep-th].
  %%CITATION = ARXIV:0712.0571;%%
  
  \bibitem{Guth:2007ng} 
  A.~H.~Guth,
  %``Eternal inflation and its implications,''
  J.\ Phys.\ A A {\bf 40}, 6811 (2007)
  [hep-th/0702178 [HEP-TH]].
  %%CITATION = HEP-TH/0702178;%%

\bibitem{Linde:2007fr} 
  A.~D.~Linde,
  %``Inflationary Cosmology,''
  Lect.\ Notes Phys.\  {\bf 738}, 1 (2008)
  [arXiv:0705.0164 [hep-th]].
  %%CITATION = ARXIV:0705.0164;%%
  
  \bibitem{CDL} 
  S.~R.~Coleman and F.~De Luccia,
  %``Gravitational Effects on and of Vacuum Decay,''
  Phys.\ Rev.\ D {\bf 21}, 3305 (1980), 
  %%CITATION = PHRVA,D21,3305;%%
  S.~R.~Coleman,
  %``The Fate of the False Vacuum. 1. Semiclassical Theory,''
  Phys.\ Rev.\ D {\bf 15}, 2929 (1977)
  [Erratum-ibid.\ D {\bf 16}, 1248 (1977)],
  %%CITATION = PHRVA,D15,2929;%%
  C.~G.~Callan, Jr. and S.~R.~Coleman,
  %``The Fate of the False Vacuum. 2. First Quantum Corrections,''
  Phys.\ Rev.\ D {\bf 16}, 1762 (1977).
  %%CITATION = PHRVA,D16,1762;%%
  
\bibitem{Garriga:2006hw}
  J.~Garriga, A.~H.~Guth, A.~Vilenkin,
  %``Eternal inflation, bubble collisions, and the persistence of memory,''
  Phys.\ Rev.\  {\bf D76}, 123512 (2007).
  [hep-th/0612242].
  
  \bibitem{Aguirre:2007an}
  A.~Aguirre, M.~CJohnson, A.~Shomer,
  %``Towards observable signatures of other bubble universes,''
  Phys.\ Rev.\  {\bf D76}, 063509 (2007).
  [arXiv:0704.3473 [hep-th]].
  
  \bibitem{Aguirre:2007wm}
  A.~Aguirre, M.~CJohnson,
  %``Towards observable signatures of other bubble universes. II: Exact solutions for thin-wall bubble collisions,''
  Phys.\ Rev.\  {\bf D77}, 123536 (2008).
  [arXiv:0712.3038 [hep-th]].
  
  \bibitem{Aguirre:2008wy}
  A.~Aguirre, M.~C.~Johnson, M.~Tysanner,
  %``Surviving the crash: assessing the aftermath of cosmic bubble collisions,''
  Phys.\ Rev.\  {\bf D79}, 123514 (2009).
  [arXiv:0811.0866 [hep-th]].
  
  \bibitem{Chang:2007eq} 
  S.~Chang, M.~Kleban and T.~S.~Levi,
  %``When worlds collide,''
  JCAP {\bf 0804}, 034 (2008)
  [arXiv:0712.2261 [hep-th]].
  %%CITATION = ARXIV:0712.2261;%%

  \bibitem{Chang:2008gj}
  S.~Chang, M.~Kleban and T.~S.~Levi,
  %``Watching Worlds Collide: Effects on the CMB from Cosmological Bubble
  %Collisions,''
  JCAP {\bf 0904}, 025 (2009)
  [arXiv:0810.5128 [hep-th]].
  %%CITATION = JCAPA,0904,025;%%
  
  \bibitem{Freivogel:2009it}
  B.~Freivogel, M.~Kleban, A.~Nicolis, K.~Sigurdson,
  %``Eternal Inflation, Bubble Collisions, and the Disintegration of the Persistence of Memory,''
  JCAP {\bf 0908}, 036 (2009).
  [arXiv:0901.0007 [hep-th]].
  
  \bibitem{Larjo:2009mt} 
  K.~Larjo and T.~S.~Levi,
  %``Bubble, Bubble, Flow and Hubble: Large Scale Galaxy Flow from Cosmological Bubble Collisions,''
  JCAP {\bf 1008}, 034 (2010)
  [arXiv:0910.4159 [hep-th]].
  %%CITATION = ARXIV:0910.4159;%%
  
  \bibitem{Czech:2010rg}
  B.~Czech, M.~Kleban, K.~Larjo, T.~S.~Levi and K.~Sigurdson,
  %``Polarizing Bubble Collisions,''
  JCAP {\bf 1012}, 023 (2010)
  [arXiv:1006.0832 [astro-ph.CO]].
  %%CITATION = JCAPA,1012,023;%%
  
  \bibitem{Easther:2009ft} 
  R.~Easther, J.~T.~Giblin, Jr, L.~Hui and E.~A.~Lim,
  %``A New Mechanism for Bubble Nucleation: Classical Transitions,''
  Phys.\ Rev.\ D {\bf 80}, 123519 (2009)
  [arXiv:0907.3234 [hep-th]].
  %%CITATION = ARXIV:0907.3234;%%
  
  \bibitem{Giblin:2010bd} 
  J.~T.~Giblin, Jr, L.~Hui, E.~A.~Lim and I-S.~Yang,
  %``How to Run Through Walls: Dynamics of Bubble and Soliton Collisions,''
  Phys.\ Rev.\ D {\bf 82}, 045019 (2010)
  [arXiv:1005.3493 [hep-th]].
  %%CITATION = ARXIV:1005.3493;%%
  
 \bibitem{Johnson:2010bn} 
  M.~C.~Johnson and I-S.~Yang,
  %``Escaping the crunch: Gravitational effects in classical transitions,''
  Phys.\ Rev.\ D {\bf 82}, 065023 (2010)
  [arXiv:1005.3506 [hep-th]].
  %%CITATION = ARXIV:1005.3506;%%
  
  \bibitem{Kleban:2011yc} 
  M.~Kleban, T.~S.~Levi and K.~Sigurdson,
  %``Observing the Multiverse with Cosmic Wakes,''
  arXiv:1109.3473 [astro-ph.CO].
  %%CITATION = ARXIV:1109.3473;%%
  
\bibitem{Gobbetti:2012yq} 
  R.~Gobbetti and M.~Kleban,
  %``Analyzing Cosmic Bubble Collisions,''
  arXiv:1201.6380 [hep-th].
  %%CITATION = ARXIV:1201.6380;%%
  
\bibitem{Feeney:2010dd}
  S.~M.~Feeney, M.~C.~Johnson, D.~J.~Mortlock, H.~V.~Peiris,
  %``First Observational Tests of Eternal Inflation: Analysis Methods and WMAP 7-Year Results,''
  
  [arXiv:1012.3667 [astro-ph.CO]].
  
  \bibitem{McEwen:2012uk} 
  J.~D.~McEwen, S.~M.~Feeney, M.~C.~Johnson and H.~V.~Peiris,
  %``Optimal filters for detecting cosmic bubble collisions,''
  arXiv:1202.2861 [astro-ph.CO].
  %%CITATION = ARXIV:1202.2861;%%
  
  \bibitem{Johnson:2011wt} 
  M.~C.~Johnson, H.~V.~Peiris and L.~Lehner,
  %``Determining the outcome of cosmic bubble collisions in full General Relativity,''
  arXiv:1112.4487 [hep-th].
  %%CITATION = ARXIV:1112.4487;%%
  
  \bibitem{Hwang:2012pj} 
  D.~-i.~Hwang, B.~-H.~Lee, W.~Lee and D.~-h.~Yeom,
  %``Bubble collision with gravitation,''
  arXiv:1201.6109 [gr-qc].
  %%CITATION = ARXIV:1201.6109;%%
  
\bibitem{Aguirre:2009ug}
  A.~Aguirre and M.~C.~Johnson,
  %``A Status report on the observability of cosmic bubble collisions,''
  Rept.\ Prog.\ Phys.\  {\bf 74}, 074901 (2011)
  [arXiv:0908.4105 [hep-th]].
  %%CITATION = RPPHA,74,074901;%%
  
  \bibitem{Kleban:2011pg}
  M.~Kleban,
  %``Cosmic Bubble Collisions,''
  [arXiv:1107.2593 [astro-ph.CO]].
  
  \bibitem{WMAP2} 
  E.~Komatsu {\it et al.}  [WMAP Collaboration],
  %``Seven-Year Wilkinson Microwave Anisotropy Probe (WMAP) Observations: Cosmological Interpretation,''
  Astrophys.\ J.\ Suppl.\  {\bf 192}, 18 (2011)
  [arXiv:1001.4538 [astro-ph.CO]].
  %%CITATION = ARXIV:1001.4538;%%
  
  \bibitem{Sachs:1967er} 
  R.~K.~Sachs and A.~M.~Wolfe,
  %``Perturbations of a cosmological model and angular variations of the microwave background,''
  Astrophys.\ J.\  {\bf 147}, 73 (1967)
  [Gen.\ Rel.\ Grav.\  {\bf 39}, 1929 (2007)].
  %%CITATION = ASJOA,147,73;%%
   
 
  \bibitem{Erickcek:2008jp}
  A.~L.~Erickcek, S.~M.~Carroll and M.~Kamionkowski,
  %``Superhorizon Perturbations and the Cosmic Microwave Background,''
  Phys.\ Rev.\  D {\bf 78}, 083012 (2008)
  [arXiv:0808.1570 [astro-ph]];
  %%CITATION = PHRVA,D78,083012;%%
  J.~P.~Zibin and D.~Scott,
  %``Gauging the cosmic microwave background,''
  Phys.\ Rev.\ D {\bf 78}, 123529 (2008)
  [arXiv:0808.2047 [astro-ph]].
  %%CITATION = ARXIV:0808.2047;%%
  
  \bibitem{Guth:2012ww} 
  A.~H.~Guth and Y.~Nomura,
  %``What can the observation of nonzero curvature tell us?,''
  arXiv:1203.6876 [hep-th].
  %%CITATION = ARXIV:1203.6876;%%
  
  \bibitem{Gordon:2005ai}
  C.~Gordon, W.~Hu, D.~Huterer, T.~M.~Crawford,
  %``Spontaneous isotropy breaking: a mechanism for cmb multipole alignments,''
  Phys.\ Rev.\  {\bf D72}, 103002 (2005).
  [astro-ph/0509301].
  
  \bibitem{book}
  D.~A.~Varshalovich, A.~N.~Moskalev, and V.~K.~Khersonskii,
  {\it Quantum Theory of Angular Momentum} (World Scientific, 1988).
  
\bibitem{Doran:2003sy}
  M.~Doran,
  %``Cmbeasy:: an object oriented code for the cosmic microwave background,''
  JCAP {\bf 0510}, 011 (2005).
  [astro-ph/0302138].

  \bibitem{WMAP} 
  N.~Jarosik, C.~L.~Bennett, J.~Dunkley, B.~Gold, M.~R.~Greason, M.~Halpern, R.~S.~Hill and G.~Hinshaw {\it et al.},
  %``Seven-Year Wilkinson Microwave Anisotropy Probe (WMAP) Observations: Sky Maps, Systematic Errors, and Basic Results,''
  Astrophys.\ J.\ Suppl.\  {\bf 192}, 14 (2011)
  [arXiv:1001.4744 [astro-ph.CO]].
  %%CITATION = ARXIV:1001.4744;%%
  
  \bibitem{WMAP_quad} 
  C.~L.~Bennett, R.~S.~Hill, G.~Hinshaw, D.~Larson, K.~M.~Smith, J.~Dunkley, B.~Gold and M.~Halpern {\it et al.},
  %``Seven-Year Wilkinson Microwave Anisotropy Probe (WMAP) Observations: Are There Cosmic Microwave Background Anomalies?,''
  Astrophys.\ J.\ Suppl.\  {\bf 192}, 17 (2011)
  [arXiv:1001.4758 [astro-ph.CO]].
  %%CITATION = ARXIV:1001.4758;%%
  
  \bibitem{Aguirre:2006ak} 
  A.~Aguirre, S.~Gratton and M.~CJohnson,
  %``Hurdles for recent measures in eternal inflation,''
  Phys.\ Rev.\ D {\bf 75}, 123501 (2007)
  [hep-th/0611221].
  %%CITATION = HEP-TH/0611221;%%
  
 \end{thebibliography}
\end{document}